\documentstyle[epsf]{elsart}

\newcommand{\mlabel}[1]{\label{#1}}
\newcommand{\mref}[1]{(\ref{#1})}

\newcommand{\intsub}[1]{   \!\!\!\!\!\!\!\!\!\!
        \int\limits_{
          {}^{ \phantom{D}}
          _{\scriptstyle #1 } }
          \!\!\!\!\!\!\!\!\!\!\!
          }
\newcommand{\mal}{\cdot}
\newcommand{\third}{\frac13}
\newcommand{\eq}[1]{Eq.~\mref{#1}}
\newcommand{\eqs}[1]{Eqs.\ \mref{#1}}
\newcommand{\projector}{{\cal P}}
\newcommand{\rf}[1]{Ref.~\cite{#1}}

\newcommand{\kapitel}[1]{Sec.~\ref{#1}}
\newcommand{\vA}{\vec{A}}
\newcommand{\vD}{\vec{D}}
\newcommand{\vE}{\vec{E}}

\newcommand{\va}{\vec{a}}
\newcommand{\ve}{\vec{e}}

\newcommand{\vk}{\vec{k}}
\newcommand{\vp}{\vec{p}}

\newcommand{\vv}{\vec{v}}
\newcommand{\vx}{\vec{x}}
\newcommand{\lsi}{\raise0.3ex\hbox{$<$\kern-0.75em\raise-1.1ex\hbox{$\sim$}}}
\newcommand{\gsi}{\raise0.3ex\hbox{$>$\kern-0.75em\raise-1.1ex\hbox{$\sim$}}}
\newcommand{\lsim}{\mathop{\lsi}}
\newcommand{\gsim}{\mathop{\gsi}}
\renewcommand{\vec}[1]{{\bf #1}}
\newcommand{\stern}{}

\renewcommand{\(}{\left(}
\renewcommand{\)}{\right)}
\renewcommand{\[}{\left[}
\renewcommand{\]}{\right]}

\newcommand{\mmdebye}{m^2_{\rm D}}
\newcommand{\mdebye}{{m_{\rm D}}}
\newcommand{\tr}{{\rm t}}

\renewcommand{\in}{{\rm in}}
\newcommand{\im}{{\rm im}}

\newcommand{\intkd}[1]{\int\frac{d^3 k_{#1}}{(2\pi)^3}}
\newcommand{\intv}[1]{\int\frac{d\Omega_{{\rm v}_{#1}}}{4\pi}}

\newcommand{\deltav}{\delta^{(S^2)}}
\newcommand{\lav}{\langle\!\langle}
\newcommand{\rav}{\rangle\!\rangle}
\newcommand{\nn}{\nonumber}
\renewcommand{\theequation}{\thesection.\arabic{equation}}
\newcommand{\msection}[1]{\section{#1}\setcounter{equation}0}

\begin{document}
\begin{flushright}
NBI-HE-99-13\\
\end{flushright}
\begin{frontmatter}
\title{\bf From hard thermal loops  to  Langevin dynamics}
\author{Dietrich B\"odeker\thanksref{email}}
\thanks[email]{e-mail: bodeker@nbi.dk}

\address{The Niels Bohr Institute,
Blegdamsvej 17, DK-2100 Copenhagen \O, Denmark}

\begin{abstract} 
In hot non-Abelian gauge theories, processes characterized by the
momentum scale $g^2 T$ (such as electroweak baryon number violation in
the very early universe) are non-perturbative.  An effective theory
for the soft ($|\vec{p}|\sim g^2 T$) field modes is obtained by
integrating out momenta larger than $g^2 T$. Starting from the hard
thermal loop effective theory, which is the result of integrating out
the scale $T$, it is shown how to integrate out the scale $gT$ in an
expansion in the gauge coupling $g$.  At leading order in $g$, one
obtains Vlasov-Boltzmann equations for the soft field modes, which
contain a Gaussian noise and a collision term. The 2-point function
of the noise and the collision term are explicitly calculated in a
leading logarithmic approximation. In this approximation the Boltzmann
equation is solved. The resulting effective theory for the soft field
modes is described by a Langevin equation. It determines the
parametric form of the hot baryon number violation rate as $\Gamma =
\kappa g^{10} \log(1/g) T^4$, and it allows for a calculation of
$\kappa$ on the lattice.
\end{abstract}
\end{frontmatter}

\msection{Introduction}\mlabel{sec.introduction}
Due to the chiral anomaly, baryon number is not conserved in the
standard electroweak theory \cite{thooft}. At zero temperature, the
rate for baryon number changing processes is exponentially small. It
becomes unsuppressed when $T$ is of order of the electroweak phase
transition or cross-over temperature $T_{\rm c}\sim 100$~GeV
\cite{krs}. Knowledge of this rate is a prerequisite for any attempt
to understand the baryon asymmetry of the universe \cite{rubakov}.

Baryon number changing processes are due to topology changing
transitions of the weak SU(2) gauge fields. In the high temperature
phase ($T\gsim T_{\rm c}$) these transitions are unsuppressed only
when the participating gauge fields are transverse and have a
wavelength of order $(g^2 T)^{-1}$ \cite{arnoldmclerran}, where $g$ is
the weak gauge coupling. This is the scale at which finite temperature
perturbation theory for non-Abelian gauge theories breaks down
\cite{linde80,gross}. Therefore, the rate in the high temperature
phase, the so-called hot sphaleron rate, cannot be computed using
weak coupling methods.

Close to thermal equilibrium, the rate for topology changing
transitions can be written in terms of a different-time correlation
function of the form
\begin{eqnarray}
        \mlabel{c}
        C(t_1 - t_2) = \langle {\cal O}(t_1) {\cal O}(t_2)\rangle
	,
\end{eqnarray}
where $\langle \cdots \rangle$ denotes the average over a thermal
ensemble \cite{khlebnikov}. The operator ${\cal O}(t)$ is a gauge
invariant function of the gauge fields $A_\mu(t,\vx)$ at time
$t$. Computing such correlation functions non-perturbatively is a
difficult task.  Euclidean lattice simulation are of no use because
this is a problem with real (Minkowski) time.

In this paper it is shown that correlation functions like \mref{c} are
determined by an effective classical field theory which is described
by \eqs{langevin2} and \mref{langevintag}. The great advantage
classical field theories have over quantum field theories at finite
temperature is that one can treat them non-perturbatively, like, e.g.
on a lattice. The use of classical field theory for computing the hot
sphaleron rate was suggested more than ten years ago
\cite{grigoriev}. The ``soft'' field modes, i.e. modes with spatial
momenta of order $g^2 T$, are classical because they contain a large
number of quanta as can be estimated from the Bose distribution
function\footnote{For spatial vectors I use the notation
$p=|\vp|$. Four-vectors are denoted by capitals, $P^\mu = (p_0,\vp)$
and I use the metric $P^2 = p_0^2 - p^2$.}
\begin{eqnarray} 
	n(p)= \frac{1}{e^{p/T} - 1} \simeq \frac{T}{p} \gg 1 \qquad
	(p\ll T) .
\end{eqnarray}
However, not all field modes which are relevant to the problem are
classical.  The ``hard'' ($p\sim T$) modes have occupation number of
order 1 and they strongly affect the non-perturbative gauge field
dynamics by Landau damping which was realized by Arnold, Son, and
Yaffe \cite{asy}.

In order to be able to use the classical field approximation, one has
to integrate out the hard modes which can be done in perturbation
theory. At leading order one obtains the well known hard thermal loops
\cite{pisarski}. Time independent problems can be described in the
imaginary time formalism. Then, the only effect of hard thermal loops
is the Debye screening of electric interactions on the length scale
$(gT)^{-1}$. For time dependent problems hard thermal loops affect
both electric (longitudinal) and magnetic (transverse) degrees of
freedom. The Landau damping (or dynamical screening) is described by a
discontinuity of the hard thermal loop resummed propagator for
space-like momenta. In \rf{letter}, it was realized that hard thermal
loop induced interactions between soft and ``semi-hard'' ($p\sim gT$)
modes have an effect on the soft dynamics which is even stronger than
Landau damping by a logarithm of $1/g$. These interactions
determine the characteristic time scale associated with the soft
non-perturbative dynamics as
\begin{eqnarray}
	t\sim \frac{1}{g^4 T \log(1/g)}
	.
\end{eqnarray}

The hard thermal loop effective theory describes field modes with
$p\ll T$ and should therefore be classical. Nevertheless, it is
difficult to use for non-perturbative lattice calculations. Like any
classical field theory at finite temperature it is plagued with
Rayleigh-Jeans UV divergences which are the same as in classical
finite temperature Yang-Mills theory. For equal-time correlation
functions the hard thermal loop effective theory reduces to a
3-dimensional Euclidean field theory which corresponds to dimensional
reduction at lowest order. In this case, the divergences can be
removed by local counter-terms. For different-time correlation
functions this is not possible. There are linear divergences which are
non-local in space and time and on the lattice they even break
rotational invariance \cite{bms,arnold}.\footnote{This problem does not arise
in scalar $\phi^4$ theory \cite{aarts,wb} where the ultraviolet
divergences in different time correlation functions are the same as in
dimensional reduction.}  This is in sharp contrast with Euclidean
lattice theories, where rotational non-invariant terms vanish in the
limit of small lattice spacing.\footnote{There are proposals for using
the hard thermal loop effective theory for lattice simulations
\cite{bms,hu1}. In \cite{hu2}, the method suggested in \cite{hu1} was
used to compute the hot sphaleron rate~\cite{hu2} and the sphaleron
rate in the symmetry broken phase~\cite{moorebroken}. One has to use a
large numerical value of the Debye mass such that the physical hard
thermal loops dominate over the UV divergences. This means that one
cannot take the continuum limit at fixed Debye mass.}

Fortunately, one can also integrate out the momentum scale $gT$ in
perturbation theory. In this way one obtains an effective theory for
the soft field modes only. Since the dynamics of the soft modes is
strongly damped, this effective theory has a much better UV behaviour
than classical Yang-Mills theory. In a leading logarithmic
approximation \cite{letter} it is even UV finite \cite{asy2}. It has
recently been used for a lattice calculation of the hot sphaleron rate
by Moore~\cite{moorelog}.

The most direct approach to integrating out the scale $gT$ is to
compute loop diagrams within the hard thermal loop effective theory
with soft external momenta and semi-hard loop momenta.  This has been
done at one loop level in Ref.\ \cite{ladder}.  However, the one loop
approximation is not sufficient for obtaining the correct effective
theory for the soft field modes.  Computing higher loop diagrams in
the hard thermal loop effective theory is quite tedious. It turns out
to be much more efficient to use the formulation of hard thermal loop
effective theory in terms of kinetic equations, which is the subject
of this paper.

The main results of this paper and a brief description of the method
employed have been presented in \rf{letter} (for a brief review, see
\cite{regensburg}). Here, the calculation is described in detail with
emphasis on the accuracy of the various approximations. An argument
which was used in \cite{letter} when solving the (leading log)
Boltzmann equation \mref{boltzmann.2} was not rigorous. This is
improved upon in \kapitel{sec.solving.boltzmann}, while the result of
\cite{letter} is not affected. An alternative (and also rigorous)
method for solving \mref{boltzmann.2} has been presented by Arnold,
Son, and Yaffe \cite{asy3}. In \cite{asy2} these authors have
described a simple physical picture for the results of \rf{letter},
using the concept of color conductivity \cite{gyulassy,heiselberg}.
Recently, the Boltzmann equation was also obtained in
\cite{asy2},\cite{litim,valle}. A recent detailed derivation of the
Boltzmann equation can be found in \cite{blaizot99}.

The paper is organized as follows. The starting point is the hard
thermal loop effective theory for gauge field modes with momenta small
compared with the temperature. It can be conveniently formulated in
terms of kinetic equations derived by Blaizot and Iancu and
independently by Nair, which are briefly reviewed in
\kapitel{sec.htl}. In order to integrate out the scale $gT$, the
fields in the kinetic equations are decomposed into soft and semi-hard
components (\kapitel{sec.separation}). The effect of the semi-hard
fields is described by the term $\xi$ in the equations of motion for
the soft fields, \eq{vlasovsoft}. The calculation of $\xi$ is the
subject of Sects.  \ref{sec.solving.eom}-\ref{sec.xiexpansion}. First,
the equations of motion for the semi-hard fields are formally solved
in \kapitel{sec.solving.eom}. There are two types of approximations
which will be made, these are qualitatively described in
\kapitel{sec.approximations}. The first type of approximation is
discussed in \kapitel{sec.factorization} for a scalar toy model. For
the second approximation one needs to know the characteristic
amplitudes of the soft fields (\kapitel{sec.scales}). Then, the
perturbative expansion of $\xi$ is discussed in
\kapitel{sec.xiexpansion}. At leading order in $g$, the equation of
motion for the soft fields is described by Vlasov-Boltzmann equations
containing a Gaussian noise and a collision term. In
\kapitel{sec.logarithmic} the 2-point function of the noise and the
collision term are explicitly calculated in a leading logarithmic
approximation. In this approximation the Boltzmann equation is solved,
and one obtains a Langevin equation for the soft gauge fields.  It
determines the (leading log) parametric form of the hot sphaleron rate
(\kapitel{sec.sphaleron}).

\msection{Hard thermal loops}\mlabel{sec.htl}
The first step towards an effective classical theory for the soft gauge
fields is to integrate out the hard field modes. Since the hard modes
are weakly interacting, this can be done in the one-loop
approximation. The dominant contributions are obtained when
one momentum in the loop is on shell, $Q^2 =0$.\footnote{We consider
only particles with masses much smaller than the temperature.} For the
remaining propagators one can use the large energy (or eikonal)
approximation
\begin{eqnarray}
	\mlabel{eikonal.propagator}
	\frac{1}{(Q + P)^2} = \frac{1}{2 Q\cdot P+ P^2}
	\simeq \frac{1}{2 Q\cdot P}
	= \frac{1}{2 q} \frac{1}{v\cdot P}
	,	
\end{eqnarray}
where $P$ is some linear combination of the external momenta and $v =
Q/q$. In this way one obtains the so called hard thermal loops
\cite{pisarski}.

In Abelian theories, there is only a hard thermal loop 2-point
function, while there are hard thermal $n$-point functions for any $n$
in non-Abelian theories.  This point is essential for the calculation
in this paper. In \cite{letter,ladder} it was shown that at one-loop order
in the hard thermal loop effective theory the dominant contribution
to the effective theory for the soft modes is due to hard thermal loop
vertices.

The hard thermal loop effective theory is an effective theory for
momenta $p$ small compared with $T$. Therefore, it can be described by
classical equations of motion. For the present problem this leads to
an enormous simplification compared with the diagrammatic approach in
\rf{ladder}.  However, the hard thermal loop equations of motion
\cite{jackiw} are non-local, which is caused by the term $1/v\mal P$
in the eikonal propagator \mref{eikonal.propagator}. The non-locality
is due to the fact that the hard particles, which are the quanta of
the hard modes, can move undisturbed over distances as large as the
typical size of the low momentum fields.

Fortunately, there is a local formulation of the hard thermal loop
equations of motion which was obtained by Blaizot and Iancu and by
Nair
\cite{blaizot93}-\cite{Nair:hamiltonian}
(see also \cite{kelly}). It is the non-Abelian generalization of the
linearized Vlasov equations for an electro-magnetic plas\-ma (see
e.g. \cite{landau10}). In addition to the gauge fields, these
equations contain fields $W^a(x,\vv)$ describing the fluctuations of
the phase space density of hard particles; $\vv$ is the 3-velocity of
the particles with $\vv^2 = 1$. The $W^a$ transform under the adjoint
representation of the gauge group.  The first equation is the Maxwell
equation for the gauge fields
\begin{eqnarray}
        \mlabel{maxwell}
        [D_\mu, F^{\mu\nu}(x)]=  \mmdebye \int\frac{d\Omega_\vec{v}}{4\pi}
        v^\nu  W(x,\vv)
	,
\end{eqnarray}
where $D_\mu = \partial_\mu - i g A_\mu$ is the covariant derivative.
Furthermore, $A_\mu(x) = A_\mu^a(x)T^a$ with Hermitian generators of
the fundamental representation which satisfy $[T^a,T^b] = i f^{abc}
T^c$ and which are normalized such that Tr$(T^a T^b) =(1/2)
\delta^{ab}$. All fields coupling to the gauge fields contribute to
the hard thermal loops in the same functional form.  In a SU($N$)
gauge theory with $n_f$ chiral fermions (or $2n_f$ Dirac fermions) and
$N_s$ scalars in the fundamental representation, one has $\mmdebye=
(1/3) (N + N_s/2 + n_f/4)g^2 T^2$. The rhs of \mref{maxwell} is the
current due to the hard particles. The integral in Eq.\mref{maxwell}
is over the directions of the unit vector $\vv$. The equation of
motion for $W$ reads
\begin{eqnarray}
        \mlabel{vlasov}
        [v\cdot D, W(x,\vv)] = \vec{v}\cdot\vec{E} (x)
	,
\end{eqnarray}
where $\vec{E}$ is the electric field strength tensor.  The conserved
Hamiltonian corresponding to Eqs.\ \mref{maxwell}-\mref{vlasov} is
\cite{Blaizot:energy,Nair:hamiltonian}
\begin{eqnarray}
        H&=&\int d^3 x {\rm Tr} \Bigg\{
        \vec{E}(x)\cdot\vec{E}(x) +\vec{B}(x)\cdot\vec{B}(x) \nn\\ 
        \mlabel{hamiltonian} 
        &&{}\hspace{5cm}+
        \mmdebye\int\frac{d\Omega_\vec{v}}{4\pi}
        W(x,\vv) W(x,\vv)\Bigg\}
	.
\end{eqnarray}

Different-time correlation functions like \mref{c} can then be
computed as follows:
\begin{enumerate}
\item Compute the solution $A(x)$ of the non-Abelian Vlasov
  equations \mref{maxwell}-\mref{vlasov} for given initial conditions
  $A_{\rm in}(\vx)$, $\vec{E}_{\rm in}(\vx)$ and $W_{\rm in}(\vx,\vv)$
  at $t=0$.
\item Insert the solution into $ {\cal O}[A(t_i)]$.
\item Then $C(t_1 - t_2)$ is given by the product ${\cal O}[A(t_1)]
  {\cal O}[A(t_2)]$ averaged over the initial conditions with the
  Boltzmann weight $\exp(-\beta H[A_{\rm in},\vec{E}_{\rm in},W_{\rm
  in}])$.
\end{enumerate}

\msection{Separating soft and semi-hard degrees of freedom}
\mlabel{sec.separation}
The fields in the kinetic equations \mref{maxwell} and  \mref{vlasov}
contain Fourier components with momenta of order $gT$ and $g^2 T$.
The goal of this paper is to obtain equations of motion for the soft
fields only.  We introduce a separation scale $\mu$ such that
\begin{eqnarray}
	\mlabel{mu}
        g^2 T\ll \mu \ll gT. 
\end{eqnarray}
The fields $A$, $\vec{E}$ and $W$ are decomposed into soft
and semi-hard  modes,
\begin{eqnarray}
   \mlabel{separation}
   A \to  A + a ,\quad
   \vec{E} \to  \vec{E} + \vec{e} ,\quad
   W \to W + w 
	.
\end{eqnarray}
The soft fields\footnote{For notational simplicity, no new symbols are
introduced for the soft modes. From now on $A$, $\vec{E}$ and $W$ will
always refer to the soft fields only.} $A$, $\vec{E}$, and $W$ contain
the spatial Fourier components with $p<\mu$ while the semi-hard ones
$a$, $\ve$, and $w$ consist of those with $k>\mu$.  Both $W$ and $w$
describe fluctuations of the hard particle distribution. $W$ ($w$) is
the long (short) wavelength part this distribution varying on length
scale greater (less) than $1/\mu$. A word on notation: below, soft
momenta will always be denoted by $P$, while semi-hard momenta will be
referred to by $K$.

In order to obtain a gauge invariant effective theory for the soft
fields, the separation has to be done in a way which does not break
gauge invariance. In this paper the precise form of the effective
theory will be calculated explicitly only in a leading logarithmic
approximation. For this purpose it is sufficient to keep only those
terms which depend logarithmically on the separation scale. For these
terms it is irrelevant how the cutoff is realized exactly.

In the following, we will obtain two sets of kinetic equations, one
for the soft fields and one for the semi-hard ones.  Due to the
non-linear terms these are coupled. Then, the semi-hard fields
will be eliminated and one obtains equations of motion containing only
the soft fields.

After the decomposition \mref{separation}, the field
strength tensor becomes
\begin{eqnarray}
        \mlabel{F.0.3.2}
        F^{\mu\nu}\to
        F^{\mu\nu} + f^{\mu\nu} - i g[a^\mu,a^\nu]
	,
\end{eqnarray}
where 
\begin{eqnarray}
        \mlabel{F.0.3.3}
        f^{\mu\nu} = [D^\mu, a^\nu] - [D^\nu, a^\mu]
	.
\end{eqnarray}
The term $F^{\mu\nu}$ on the rhs of \eq{F.0.3.2} and the covariant
derivatives in \eq{F.0.3.3} contain only the soft gauge fields.
The lhs of the Maxwell equation \mref{maxwell} becomes
\begin{eqnarray}
        \nn 
        [D_\mu,F^{\mu\nu}] &\to& [D_\mu,F^{\mu\nu}] + [D_\mu,f^{\mu\nu}] 
        - i g [a_\mu, F^{\mu\nu}]
\\      
        \mlabel{F.0.4.1}        
        &&
        -i g  [ D_\mu,[a^\mu, a^\nu]] 
         -ig [a_\mu, f^{\mu\nu}]  - g^2[ a_\mu,[a^\mu, a^\nu]]
	.
\end{eqnarray}

\begin{figure}[t]
 
 
\hspace{.8cm}
\epsfysize=6.5cm
\centerline{
        \hspace{-.3cm}
        \epsfysize=7cm\epsffile{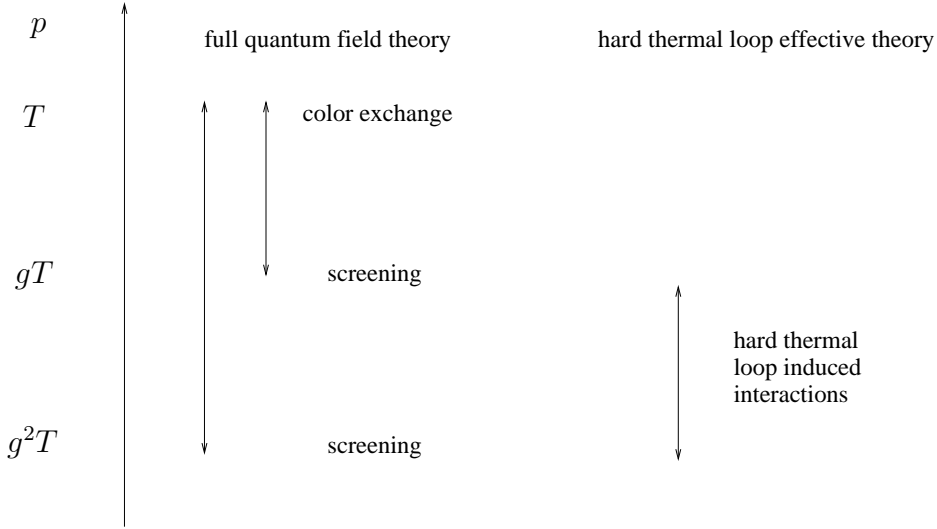}
        }

\vspace*{-2.5cm}
\begin{picture}(300,100)
\put(14,218){$p$}
\put(11,182){$T$}
\put(8,123){$g T$}
\put(5,59){$g^2 T$}
\end{picture}

\caption[a]{The interactions which are relevant to the leading order
effective theory for the soft gauge fields. The hard modes dynamically
screen the soft and semi-hard fields. The relevant hard--semi-hard interaction
is small angle scattering which exchanges color charge between hard
particles (see \kapitel{sec.logarithmic}). Soft--semi-hard
interactions in the full quantum field theory can be
neglected. Therefore, the relevant interactions within the hard
thermal loop effective theory are due to hard thermal loop vertices.}
\mlabel{fig.interactions}
\vspace*{.5cm}
\end{figure}

Inspecting one-loop diagrams in the hard thermal loop effective theory
\cite{ladder} one can see that in order to integrate out the
semi-hard modes only hard thermal loop vertices are relevant at
leading order and that diagrams containing tree level vertices are
suppressed by a factor $g$ (see Fig.~\ref{fig.interactions}). In the
present calculation tree level vertices correspond to interaction
terms in \eq{F.0.4.1}. Therefore, at leading order in $g$, one can
drop all interaction terms in \mref{F.0.4.1} which contain semi-hard
fields, i.e. all but the first two terms. In addition, the covariant
derivatives in the second term can be replaced by ordinary
derivatives.

The lhs of \eq{vlasov} becomes
\begin{eqnarray}
        \mlabel{F.0.5.1}        
        [v\mal D, W] \to [v\mal D, W] + [v\mal D, w] 
        - ig [v\mal a, W] - ig [v\mal a, w] 
	.
\end{eqnarray}
The last term on the rhs contains a product of two semi-hard fields.
Therefore, it contains both semi-hard and soft Fourier components. The
latter will turn out to be essential for the equations of motion for
the soft fields.

Assuming that the scales $gT$ and $g^2 T$ are well separated, the
decomposition \mref{separation} is invariant under the infinitesimal
gauge transformation
\begin{eqnarray}
        \mlabel{F.0.4.2}
        \delta A^\mu = [D^\mu, \omega]
\end{eqnarray}
when the gauge parameter $\omega(x) $ has only soft Fourier
components. Under \mref{F.0.4.2}, the other fields transform in the
adjoint representation,
\begin{eqnarray}
        \mlabel{F.0.4.3}
        \delta W = -ig [W,\omega]  
	,\qquad
        \delta a = -ig [a,\omega]
	,\qquad
        \delta w = -ig [w,\omega]
	.
\end{eqnarray}

\subsection{The equations of motion for the soft fields}
\mlabel{sec.soft.eom}
Dropping the self-interactions of the semi-hard fields, the soft
momentum part of the Maxwell equation \mref{maxwell} becomes
\begin{eqnarray}
        \mlabel{maxwellsoft}
        [D_\mu, F^{\mu\nu}(x)]=  \mmdebye \int\frac{d\Omega_\vec{v}}{4\pi}
        v^\nu  W(x,\vv)
	.
\end{eqnarray}
The semi-hard fields enter the equation of motion for $W$, 
\begin{eqnarray}
        \mlabel{vlasovsoft}
        [v \cdot D, W(x,\vv)]   = \vec{v}\cdot\vec{E} (x)
        + \xi(x,\vv)
	,
\end{eqnarray}
where $\xi$ is defined as 
\begin{eqnarray}
        \mlabel{F.0.5.1.5}
        \xi (x,\vv) = i g [v\mal a(x), w(x,\vv)]_{\rm soft}
	.
\end{eqnarray}
The subscript ``soft'' indicates that only spatial Fourier components
with $p<\mu$ are included.  Under the gauge transformation
\mref{F.0.4.2}, $\xi$ transforms as the fields in \eq{F.0.4.3}. Therefore,
\eqs{maxwellsoft} and \mref{vlasovsoft} are covariant under \mref{F.0.4.2}.

\subsection{The equations of motion for the semi-hard fields}
\mlabel{sec.hard.eom}
Since the semi-hard fields are perturbative, one can neglect their
self-inter\-actions. Then, the equations of motion for the semi-hard
fields become linear in $a$ and $w$,
\begin{eqnarray}
        \mlabel{F.1.1.2}
        \partial^2 a^\nu - \partial^\nu \partial_\mu a^\mu
        &=& 
         \mmdebye \int\frac{d\Omega_\vec{v}}{4\pi}
        v^\nu  w
	,
\end{eqnarray}
\begin{eqnarray}
        \mlabel{vlasovhard}
        [v\mal D, w] - i g [v\mal a, W] &=&
         \vv\mal\ve    
	.
\end{eqnarray}
In the next Section these equations will formally be solved. For that
purpose it is necessary to separate the free parts in
\mref{vlasovhard} from interaction terms. We rewrite \mref{vlasovhard}
as
\begin{eqnarray}
        \mlabel{F.1.1.6}
        v\mal\partial  w     &=& \vv\mal\ve  + h
	,
\end{eqnarray}
where the interaction terms have been collected  in 
\begin{eqnarray}
        \mlabel{F.1.1.5}
        h = ig \Big\{ [v\mal a, W] + [v\mal A, w] \Big\}
	.
\end{eqnarray}
\eqs{F.1.1.2} and \mref{F.1.1.6} will be used to write $\xi$ in terms
of the soft fields only. In \mref{F.1.1.2}, \mref{F.1.1.6} covariance
under the gauge transformation \mref{F.0.4.2} is not manifest. Only if
$\xi$ is consistently expanded in powers of $g$, gauge covariance will
be recovered.

\msection{``Solving'' the equations of motion for the semi-hard fields}
\mlabel{sec.solving.eom}
In this Section we will obtain an expansion of the semi-hard fields in
powers of the soft fields starting from the equations of motion in
\kapitel{sec.hard.eom}. This expansion is only a formal solution to
the equations of motion since the soft fields themselves depend on the
semi-hard ones. The use of this expansion will become apparent in the
\kapitel{sec.factorization}. Our method will first be illustrated with
a scalar toy model. The extension to the gauge theory case will be
straightforward.  We will restrict ourselves to the transverse
semi-hard fields because the longitudinal fields do not contribute in
the leading logarithmic approximation in \kapitel{sec.logarithmic}.

Consider a scalar field which is split into soft and semi-hard Fourier
components $\Phi$ and $\phi$. For the equation of motion for $\phi$,
corresponding to \mref{F.1.1.2} and \mref{F.1.1.6}, we write
\begin{eqnarray}
	\mlabel{F.2.1}
        {\cal D} \phi(x) = g \phi(x) \Phi(x)
	,
\end{eqnarray}
where ${\cal D} $ is some linear differential operator acting on $x$.
We will also assume that $\phi$ enters the equation of motion for
$\Phi$ in analogy with \mref{vlasovsoft}. The form of the equation of
motion for $\Phi$ is not relevant to the present discussion.

\eq{F.2.1} is equivalent to
\begin{eqnarray}
        \mlabel{F.2.2}
        \phi(x) = \phi_0(x) + g \int d^4 x' \Delta(x-x') \phi(x') \Phi(x')
	,
\end{eqnarray}
where $\phi_0$ is a solution to the free equation of motion ${\cal
D}\phi_0 =0$. The propagator $\Delta$ satisfies ${\cal D} \Delta(x-
x') =\delta^4(x-x')$. Both $\phi_0$ and the propagator $\Delta$ are
fixed by imposing initial conditions for $\phi$. We will work in
momentum space,
where \eq{F.2.2} reads
\begin{eqnarray}
        \mlabel{F.2.3}
        \phi(K) = \phi_0(K) + g \Delta(K) 
        \int \frac{d^4 p}{(2\pi)^4}  \phi(K - P) \Phi(P)
	,
\end{eqnarray}
where $k\sim  g T$ and $p \sim g^2 T$.
By iterating \eq{F.2.3} one obtains a series
\begin{eqnarray}
        \mlabel{F.2.4}
        \phi = \phi_0 + \phi_1 + \phi_2 + \cdots
	,
\end{eqnarray}
where 
\begin{eqnarray}
        \mlabel{F.17.1}
        \phi_n(K) = g \Delta (K) \int \frac{d^4 p}{(2\pi)^4}
         \phi_{n-1}(K-P)\Phi(P) 
	.
\end{eqnarray}
Each $\phi_n$ is linear in $\phi_0$. Furthermore, $\phi_n$ contains
$n$ powers of $g \Phi$.  If the soft field $\Phi$ was a fixed
background field, \eq{F.2.4} would be a solution to the equation of
motion. However, this is not the case here. $\Phi$ itself depends on
$\phi$ because the latter enters the equation of motion for $\Phi$.

The gauge theory case can be treated in full analogy with the above
example. We consider only the transverse gauge fields.  In the leading
logarithmic approximation of \kapitel{sec.logarithmic} only these will
be required. In spatial momentum space, the transverse field is
defined as
\begin{eqnarray}
	a^i_\tr(t,\vk) = \projector_\tr^{ij}(\vk) a^j(t,\vk)
	,
\end{eqnarray}
where the transverse projector is
\begin{eqnarray}
  \mlabel{projector}
  {\cal P}^{ij}_{\rm t}(\vk) = \delta^{ij} - \frac{k^i k^j}{k^2} 
	.
\end{eqnarray}
For notational simplicity, the semi-hard gauge fields will be written
without the subscript ``$\tr$'' in the following. The equations of
motion in spatial momentum space are then
\begin{eqnarray}
        \mlabel{maxwellhard.transverse}
        \ddot{\va}(t,\vk) + k^2 \va (t,\vk) =
         \mmdebye \int\frac{d\Omega_\vv}{4\pi}
        \vv_{\tr} w(t,\vk,\vv) 
	,
\end{eqnarray}
where $v^i_{\tr} = {\cal P}^{ij}_{\rm t}(\vk)v^j$. The equation for $w$
reads
\begin{eqnarray}
        \mlabel{vlasovhard.transverse}
        \dot{w} (t,\vk,\vv) + i \vv\cdot\vk w(t,\vk,\vv) =
        - \vv\cdot\dot{\va}  (t,\vk) +   h (t,\vk,\vv)
	.
\end{eqnarray}

The equations of motion can be solved by Laplace
transformation.\footnote{Or one sided Fourier transformation, cf.\
Ref.~\cite{landau10}.}  The Laplace transform of a function $f(t)$ is
\begin{eqnarray}
        \mlabel{laplace}        
        f(k^0) = \int_0^\infty dt e^{i k^0 t} f(t)
	.
\end{eqnarray}
It defines an analytic function in the upper half of the complex
$k^0$-plane.  Taking the Laplace transform of
Eqs.~\mref{maxwellhard.transverse} and \mref{vlasovhard.transverse}
one obtains (cf.\ Appendix~A)
\begin{eqnarray}
        \mlabel{maxwellhardl}
         \ve_{\rm in}(\vk) + i k_0 \va_{\rm in}(\vk) 
        - K^2 \va(K) &=&  \mmdebye \int\frac{d\Omega_\vv}{4\pi}
        \vv_{\tr} w(K,\vv) 
	,
\end{eqnarray}
\begin{eqnarray}
        \mlabel{vlasovhardl}
        -w_{\rm in}(\vk,\vv) - i v\cdot  K w(K,\vv) &=&
        \vv\cdot\va_{\rm in} (\vk) + i k^0 \vv\cdot\va(K)
        + h(K,\vv)
	,
\end{eqnarray}
where the subscript ``in'' denotes the initial values at $t=0$.
\eqs{maxwellhardl} and \mref{vlasovhardl} can be ``solved'' analogous
to \eq{F.2.3}.  For this purpose, it is convenient to write them in
the form
\begin{eqnarray}
  \mlabel{linear.a}
  -K^2 a^{i}(K) - \mmdebye \intv{}v_\tr^i w(K,\vv) &=& \ell_1^{i}(K)
	,
\end{eqnarray}
\begin{eqnarray}
  \mlabel{linear.w}
  - i k_0 \vv \mal \va(K) -i v\mal K w(K,\vv) &=& \ell_2(K,\vv)
	,
\end{eqnarray}
with 
\begin{eqnarray} 
        \mlabel{F.7.3}
        \phantom{\int}
  \ell_1^{i}(K) &=& - e_\in^i(\vk) - i k_0 a^{i}_\in(\vk)
	,
\end{eqnarray}
\begin{eqnarray}
  \mlabel{F.7.4}
  \ell_2(K,\vv) &=& w_\in(\vk,\vv) + \vv\mal\va_\in(\vk) + h(K,\vv)
	.
\end{eqnarray}
First, one solves  Eq.\ \mref{linear.w} for $w$. Inserting the
result in \mref{linear.a}, the latter can be solved for $\va$ which gives
\begin{eqnarray}
  \mlabel{solution.a.1}
  a^{i}(K) &=& \Delta_{11}^{ij}(K)\ell_1^{j}(K) 
  + \intv{}\Delta^i_{12}(K,\vv)\ell_2(K,\vv)
	.
\end{eqnarray}
Then one can determine $w$, 
\begin{eqnarray}
  \mlabel{solution.w.1} 
  w(K,\vv) &=&\Delta^i_{21}(K,\vv)\ell_1^{i}(K) 
  + \intv{1}\Delta_{22}(K,\vv,\vv_1)\ell_2(K,\vv_1)
	.
\end{eqnarray}
The propagators in \mref{solution.a.1} and \mref{solution.w.1} are
defined as
\begin{eqnarray}
        \mlabel{delta11}
        \Delta_{11}^{ij}(K) &=& \projector^{ij}_{\rm t}(\vk)
        \stern\Delta_{\rm t}(K) 
	,
\\ 
        \mlabel{delta12}
        \Delta_{12}^{i}(K,\vv) &=& 
        \mmdebye \frac{i}{v\cdot K} \stern\Delta _{\rm t}(K) v^i_\tr
	,
\\
        \mlabel{delta21}
        \Delta_{21}^{i}(K,\vv) &=& - \frac{k^0}{v\cdot K}
        \stern\Delta_{\rm t}(K)v^i_\tr
	,
\\ 
        \mlabel{delta22}
        \Delta_{22}(K,\vv,\vv') &=&
        4\pi \deltav(\vv - \vv') \frac{i}{v\cdot K} - \mmdebye
        \frac{i k^0 \vv_\tr\mal\vv'_\tr }{v\cdot K v'\cdot K}
        \stern\Delta _{\rm t}(K)
	.
\end{eqnarray}
Here $\stern\Delta_{\rm t}(K)$ denotes the hard thermal loop resummed
propagator for the transverse gauge fields
\begin{eqnarray}
        \mlabel{propagator}
        \stern\Delta_{\rm t}(K) =     
          \frac{1}{-K^2 + \delta \Pi_{\rm t}(K)}
	,
\end{eqnarray}
with the transverse hard thermal loop self-energy
\begin{eqnarray}
        \mlabel{polarization}
        \delta \Pi_{\rm t}(K) = \half
        \mmdebye k^0
        \intv{}  \frac{\vv_\tr^2}{v\cdot K}
	.
\end{eqnarray}
Furthermore, $\deltav$ is the delta-function on the two dimensional unit
sphere,
\begin{eqnarray}
        \int d\Omega_{\vv'} f(\vv') \deltav(\vv - \vv') = f(\vv) 
	.
\end{eqnarray}
Inserting \mref{F.7.3}, \mref{F.7.4} into Eqs. \mref{solution.a.1},
\mref{solution.w.1} we obtain
\begin{eqnarray}
        \mlabel{F.9.1}
        a^{i}(K) &=& 
         a^{i}_0 (K)
	+ \intv{} \Delta_{12}^i(K,\vv) h(K,\vv)
	,
\end{eqnarray}
\begin{eqnarray}
        \mlabel{F.9.2}
        w(K,\vv) &=&  w_0 (K,\vv) 
	+ \intv{1} \Delta_{22}(K,\vv,\vv_1) h(K,\vv_1)
	.
\end{eqnarray}
This is the gauge theory analogue of \eq{F.2.3}.  $\vec{a}_0$ and
$w_0$ are the solutions to the equations of motion for $h = 0$, i.e.
without the soft ``background''.  They only depend on the initial
values at $t=0$,
\begin{eqnarray}
        a_0^{i}(K) &=& 
        -\frac{i}{k^0} k^2
        \( \Delta_{11}^{ij}(K) - \Delta_{11}^{ij}(0,\vk)\)
        a_{\rm in}^{j}(\vk)  
        - \Delta_{11}^{ij}(K) e_{\rm in}^{j}(\vk) \nn
\\
        \mlabel{solutiona}
        && 
        {}+\intv{} \Delta_{12}^{i}(K,\vv) w_{\rm in}(\vk,\vv) 
	,
\end{eqnarray}
\begin{eqnarray}
        w_0(K,\vv) &=& 
        -\frac{i}{k^0} k^2
        \Delta_{21}^{i}(K,\vv)
        a_{\rm in}^{i}(\vk) - 
        \Delta_{21}^{i}(K,\vv) e_{\rm in}^{i}(\vk) \nn
\\
        \mlabel{solutionw} 
        && 
        {}+\intv{1} \Delta_{22}(K,\vv,\vv_1) w_{\rm in}(\vk,\vv_1)
	.
\end{eqnarray}
In \kapitel{sec.logarithmic} we will need the 2-point functions of
$\va_0$ and $w_0$; the results are listed in Appendix B.

By iterating Eqs.~\mref{F.9.1} and \mref{F.9.2} one obtains an
expansion
\begin{eqnarray}
        \mlabel{F.16.1.a}
        a &=& a_0 + a_1 + a_2  + \cdots
	,
\\
        \mlabel{F.16.1.w}
        w &=& w_0 + w_1 + w_2 + \cdots
	,
\end{eqnarray}
where for $n\ge 1$ 
\begin{eqnarray}
        \mlabel{F.16.2}
        a^{i}_n(K) &=&  
         \intv{} \Delta_{12}^i(K,\vv) h_n(K,\vv)
	,
\\
        \mlabel{F.16.3}
        w_n(K,\vv) &=&  
	\intv{1} \Delta_{22}(K,\vv,\vv_1) h_n(K,\vv_1)
	.
\end{eqnarray}
The term $h_n$ contains the Laplace transform of a product of
fields. Using \eq{laplace.product} it can be written as
\begin{eqnarray}
        \mlabel{F.16.5}
        h^a_n(K,\vv) &=& -g f^{abc}
        \intsub{0 < \im p^0 < \im k^0}
        \frac{d^4 p}{(2\pi)^4}  
        \[ v\cdot A^b(P) w^c_{n-1}(K-P,\vv)\nn \right. \\&& 
        ~~~~~~~~~~~~~~~~~~
                 \left. {}  -\vv\cdot\va^b_{n-1}(K-P) W^c(P,\vv)\]
	.
\end{eqnarray}
Each term in \mref{F.16.5} contains only the free solutions $\va_0$
and $w_0$ together with the full $A$ and $W$.  Since the equations of
motion \mref{F.1.1.2} and \mref{vlasovhard} are linear in $\va$ and
$w$, the $\va_n$ and $w_n$ are linear in $\va_0$ and $w_0$.

Inserting the expansion of $\va$ and $w$
into \eq{F.0.5.1.5}, one obtains an expansion for~$\xi$,
\begin{eqnarray}
        \mlabel{xiexpansion}
        \xi = \xi_0 + \xi_1 + \xi_2 + \cdots
	.
\end{eqnarray}
Each term in \mref{xiexpansion} is bilinear in the free fields $\va_0$
and $w_0$. The term $\xi_n$ is of $n$-th order in the soft
fields. Furthermore, it is of ($n$+1)-th order in the coupling $g$.
This does does not yet tell anything about the actual magnitude of $\xi_n$
which obviously depends on the amplitudes of the soft fields.

\msection{The approximations for $\xi$}\mlabel{sec.approximations}
In Sec.~\ref{sec.soft.eom} we have seen that the effect of the
semi-hard modes can be described by the term
$\xi$ in the equations of motion for the soft fields. In
\kapitel{sec.solving.eom} we have obtained an expansion of $\xi$. In the
following, it will be shown that one can make two types of
approximations for $\xi$ which both correspond to an expansion in~$g$.
\begin{enumerate}

\item The first type of approximation, which will be discussed in
\kapitel{sec.factorization}, is the factorization of the thermal
averages over initial conditions for the semi-hard fields. The 
reason why this factorization holds is that the correlation lengths
and times of the semi-hard fields are much smaller than the
corresponding ones for the soft fields. 

\item The second approximation which will be made is the truncation of
the series expansion \mref{xiexpansion}. The accuracy of this
approximation is controlled by the characteristic amplitudes of the soft
fields which will be discussed in \kapitel{sec.scales}. In
\kapitel{sec.xiexpansion} it will be shown that at leading order in
$g$ only the first two terms in \mref{xiexpansion} need to be kept.

\end{enumerate}
It should be noted that in order to go beyond leading order in $g$,
one would have to take into account terms which were already dropped
in \kapitel{sec.separation} and which are not included in the
expansion \mref{xiexpansion}. In addition, one would have to consider
corrections to the hard thermal loop approximation for the hard modes.

\msection{Factorization}\mlabel{sec.factorization}
In order to understand the approximation (i), it is instructive to go
back to the scalar field example which was discussed in the beginning
of \kapitel{sec.solving.eom}.  This will simplify the notation
significantly and the extension to the gauge theory case will again be
straightforward.  In analogy with \mref{F.0.5.1.5} we consider
\begin{eqnarray}
        \mlabel{F.17.2}
        \xi(P) = g  \int \frac{d^4 k}{(2\pi)^4} 
	\phi(K)\phi(P-K) 
	.
\end{eqnarray}
For the thermal average over initial conditions for $\phi_0$ we write
\begin{eqnarray}
        \mlabel{F.17.3}
        \langle \phi_0(K) \phi_0(K') \rangle = 
        T \tilde{\Delta}(K) (2\pi)^4\delta^4(K + K')
	,
\end{eqnarray}
instead of \mref{aa}-\mref{ww}.\footnote{In the gauge theory case, the
delta function of $k_0 + k_0'$ is the ``Laplace delta-function'' in
\eq{laplace.delta}. In this Section we will only be concerned about
the parametric form of $\xi$, for which this does not play a role.}
We will further assume that the characteristic frequency of $\phi$ is
of order $gT$.

The expansion of $\phi$ in \eq{F.17.1} yields an expansion of $\xi$
analogous to \mref{xiexpansion}. The first term in this expansion is
\begin{eqnarray}
        \mlabel{xi0.toy}
        \xi_0(P) = g  \int 
 	\frac{d^4 k}{(2\pi)^4} 
	\phi_0(K)\phi_0(P-K) 
	.
\end{eqnarray}
The next term contains one soft field and one
additional power of the coupling,
\begin{eqnarray}
        \mlabel{F.17.1.1}
        \xi_1(P) = 2 g^2 \int 
 	\frac{d^4 k}{(2\pi)^4} 
	\Delta(K) \int 
 	\frac{d^4 p_1}{(2\pi)^4} 
        \phi_0(K -P_1) \phi_0(P-K) \Phi(P_1)
	.
\end{eqnarray}

Note that in \mref{xi0.toy}, \mref{F.17.1.1}, and all higher $\xi_n$
the field $\phi_0$ always appears in the combination
\begin{eqnarray}
        \mlabel{chi}
        \chi =  \phi_0(K-P) \phi_0(P' -K)
	,
\end{eqnarray}
where $K$ is semi-hard while $P$ and $P'$ are soft. After solving the
complete equations of motion for $\phi$ and $\Phi$, the solution will
contain products of arbitrarily many factors of $\chi$. In order to
compute correlation functions like \mref{c}, one has to perform the
thermal average over initial conditions.  The key simplification which
can be made is the following: it turns out that in general, thermal
averages of products of $\chi$'s can be approximated by disconnected
parts,
\begin{eqnarray}
        \mlabel{chi.approximation}
  \langle\chi_1\cdots
  \chi_n \rangle \simeq  \langle\chi_1\rangle\cdots
  \langle\chi_n \rangle
	.
\end{eqnarray}
But since the $\chi_i$ already appear in the equation
of motion for $\Phi$, it does not make a difference
whether one first solves the complete equations of motion
and then uses \mref{chi.approximation}, or whether one replaces
\begin{eqnarray}
        \mlabel{replacechi}
        \chi \to \langle\chi\rangle
\end{eqnarray}
already in the equations of motion for $\Phi$. With the
replacement \mref{replacechi} one eliminates the semi-hard fields from
the equations of motion for the soft fields. In this way one
``integrates out'' the semi-hard fields in the present framework.

The approximation \mref{replacechi} is similar in spirit to the
derivation of kinetic equations by Klimontovich
\cite{klimontovich}.\footnote{See also \S 51 of \rf{landau10}.} The
sum of soft and semi-hard fields corresponds to ``microscopic'' and
the soft field corresponds to ``macroscopic'' degrees of freedom. The
soft fields can be considered as a spatial (not statistical) average
of the sum of soft and semi-hard fields.  Using this analogy, the
Boltzmann equation \mref{boltzmann.2} has been recently obtained in
Refs.~\cite{litim,valle}.

Now let us see why \mref{chi.approximation} holds. Let us first
estimate the thermal average of one single $\chi$,
\begin{eqnarray}
  \mlabel{F.24.2}
  \langle\chi\rangle = T 
  (2\pi)^4\delta^4(P-P') \tilde{\Delta}(K - P)
	.
\end{eqnarray}
In the final result for a non-perturbative correlation function of
soft fields, $\langle\chi\rangle$ will be integrated over the soft
momenta $P$ and $P'$. The delta-function in \mref{F.24.2} eats up one
of these integrals. Therefore, we can estimate
\begin{eqnarray}
          \mlabel{estimate.delta}
        \delta^4(P-P') \sim p_0^{-1}(g^2 T)^{-3}
	,
\end{eqnarray}
where $p_0$ is the characteristic frequency associated with $\Phi$.
In $\tilde{\Delta}(K - P)$ we can neglect the soft momentum $\vp$
relative to $\vk$ which is semi-hard.  As long as $p_0\lsim gT$, we
have $k_0-p_0\sim gT$ so that the only mass scale appearing in
$\tilde{\Delta}(K - P)$ is $gT$.  Since $\tilde{\Delta}$ has mass
dimension $-3$, we can estimate $\tilde{\Delta}(K - P)\sim (gT)^{-3}$.
In \kapitel{sec.logarithmic} we will see that the integration over
semi-hard momenta gives a logarithmic dependence on the separation
scale $\mu$. Here we will ignore these logarithms, the results of this
Section are valid beyond the leading logarithmic order considered in
\kapitel{sec.logarithmic}. Thus, we have
\begin{eqnarray}
  \mlabel{estimatechi} 
  \langle\chi\rangle \sim T  p_0^{-1} (g^2 T)^{-3} (g T)^{-3}
        \sim 
        p_0^{-1}
        g^{-9} T^{-5}
	.
\end{eqnarray}
This expression has to be compared with the connected 2-point function
of two $\chi$'s,
\begin{eqnarray}
        \langle\chi_1\chi_2
        \rangle_{\rm conn.} &=&
        T^2 (2\pi)^8
	\delta^4(P_1 + P_2 - P_1' - P_2') \tilde{\Delta}(K_1 - P_2)
        \tilde{\Delta}(P_1' - K_1)
\nn
\\
        &&
	\hspace{-1cm}
        \Big[ \delta^4(K_1 + K_2 - P_1 -P_2) +\delta^4(K_1 - K_2 - P_1  +P_2') 
        \Big]
	,
\end{eqnarray}
which contains only one delta-function of soft momenta
only. Proceeding as for Eq.~\mref{estimatechi}, we estimate
\begin{eqnarray}
  \mlabel{estimatechichi}
    \langle\chi_1\chi_2
    \rangle_{\rm conn.} 
        \sim 
         T^2 p_0^{-1} (g^2 T)^{-3} (g T)^{-10}
        \sim 
        p_0^{-1}
        g^{-16} T^{-11}
	.
\end{eqnarray}
Therefore, the connected part is suppressed relative to the
disconnected part,
\begin{eqnarray}
  \mlabel{estimatechichi.2}
    \langle\chi_1\chi_2
    \rangle_{\rm conn.} 
    \sim  \(g^2 \frac{p^0}{T}\)
    \langle\chi_1 \rangle 
    \langle \chi_2 
    \rangle
	.
\end{eqnarray}
The form of this suppression is easy to understand. The connected 
2-point function of $\chi_1\chi_2$ contains only one delta-function for
soft momenta, while the disconnected part has two. The latter has
one additional integration over semi-hard momenta $\int d^4 k
\sim(gT)^4$.  Thus, the suppression factor is
\begin{eqnarray}
        \frac{ p_0(g^2 T)^3}{(gT)^4} 
        \nn
	.
\end{eqnarray}
The same suppression occurs in higher correlation functions
\begin{eqnarray}
        \mlabel{factorization}
  \langle\chi_1\cdots
  \chi_n
    \rangle_{\rm conn.}  \sim  \(g^2 \frac{p^0}{T}\)
    \langle\chi_1 \cdots\chi_m
    \rangle_{\rm conn.}
    \langle\chi_{m+1} \cdots\chi_n
    \rangle_{\rm conn.}
	.
\end{eqnarray}
The result of this consideration is that, in general, one can
approximate
\begin{eqnarray}
        \mlabel{replacexi}
        \xi_n \simeq \lav \xi_n \rav
	,
\end{eqnarray}
where $\lav \xi_n\rav$ is defined as $\xi_n$ with $\chi$ replaced by
its thermal average.

There is one qualitative difference between the scalar model discussed
in this Section and the gauge theory. In the gauge theory case the
replacement \mref{replacechi} gives zero due to the anti-symmetry of
the structure constants. Therefore, $\xi_0$ is approximately determined
by its connected 2-point function and the discussion of $\xi_0$ will
be postponed to Sec.\ \ref{sec.xi0}.  Instead, consider the term
$\xi_1$.  After the replacement \mref{replacechi} it becomes
\begin{eqnarray}
        \mlabel{F.19.1}
        \lav\xi_1(P)\rav = 2 g^2 T \Phi(P) \int 
	\frac{d^4 k}{(2\pi)^4}
         \Delta(K)
        \tilde{\Delta}(K-P)
	.
\end{eqnarray}
Assuming that $p_0\ll gT$, one can neglect $P$ in the integral which then
becomes a dimensionless constant ($\Delta(K)$ and $\tilde{\Delta}(K)$
have mass dimension $-1$ and $-3$, respectively).  The accuracy of
this approximation is of order $g$. Then $\lav\xi_1\rav$ takes the
form
\begin{eqnarray}
        \mlabel{F.19.3}
        \lav\xi_1(P)\rav \sim  g^2T  \Phi(P)
	,
\end{eqnarray}
or,  in coordinate space,  
\begin{eqnarray}
        \mlabel{F.19.4}
        \lav\xi_1(x)\rav \sim  g^2T  \Phi(x)
	.
\end{eqnarray}
Proceeding
similarly for $n\ge 2 $, one finds that $\lav\xi_n(x)\rav$ is
proportional to $ (\Phi(x))^n$. Furthermore, $\lav\xi_n\rav$ contains
$n+1$ powers of $g$ and one factor $T$ due to \eq{F.24.2}. To get the
correct dimension, there must be $1-n $ powers of $gT$ which is the
only scale available.  Thus $ \lav\xi_n(x)\rav$ is of the form
\begin{eqnarray}
        \mlabel{F.22.1}
        \lav\xi_n(x)\rav \sim  g^n (gT)^{2-n } \(\Phi(x)\)^n
	.
\end{eqnarray}
Obviously, the expansion parameter in the expansion of $\xi$ depends
on the amplitude of the soft field (or fields in the gauge theory
case). This will be the subject of the following Section.

\msection{The characteristic amplitudes and frequencies 
of the soft fields}\mlabel{sec.scales}
In this Section we will estimate the typical amplitudes of the soft
fields, both in coordinate and in momentum space. Since the separation
\mref{separation} was defined in terms of spatial momenta, the various
soft fields can and do have different characteristic frequencies. In
\kapitel{sec.xiexpansion} these estimates will be used to determine
when the series \mref{xiexpansion} can be truncated. Strictly
speaking, some estimates can be made only a posteriori when the
effective theory for the soft dynamics is known. Then, one can go back
and check the consistency of the approximations. Some of these
estimates, together with instructive illustrations can be found in
\rf{asy}.

Any estimate for the amplitude of $\vA$ is gauge dependent. However, as
long as one works with a gauge fixing parameter which does not contain
inverse powers of the coupling, the amplitude of the soft gauge field at a
fixed time can be estimated from the tree level propagator of the
transverse gauge field which is unscreened. This gives
\begin{eqnarray}
  \mlabel{a2x}
  \left\langle \vA(t,\vx)\vA(t,\vx')\right\rangle 
	\sim \int d^3 p \frac{T}{p^2} \sim
    g^2 T^2
	.
\end{eqnarray}
Non-linear contributions to the propagator cannot be neglected, but
they do not change the order of magnitude of \mref{a2x}. Thus, from
\mref{a2x} one concludes that
\begin{eqnarray}
  \mlabel{ax}
  \vA(t,\vx)\sim gT
	.
\end{eqnarray}
Consequently, the two terms in the covariant derivative $\partial_i
-ig A_i$ are of the same order of magnitude, $\partial_i \sim g A_i$,
which makes perturbation theory for the soft gauge fields
impossible. In spatial momentum space \mref{ax} corresponds to
\begin{eqnarray}
  \mlabel{ap}
  \vA(t,\vp)\sim g^{-5} T^{-2}
	.
\end{eqnarray}

The amplitude of the field $W$ can be estimated from 
\eq{F.12.3.2}, where now the momenta are of order $g^2 T$. This gives
\begin{eqnarray}
	\langle W(t,\vp,\vv) W(t,\vp',\vv') \rangle
	&\sim& g^{-8} T^{-4}
	,
\end{eqnarray}
so that
\begin{eqnarray}
	 \mlabel{order.2.6}
	 W(t,\vp,\vv)&\sim& g^{-4} T^{-2}
	.
\end{eqnarray}
Thus, we find that in position space
\begin{eqnarray}
	 \mlabel{order.2.5}
	W(t,\vx,\vv)&\sim& g^2 T
	.
\end{eqnarray}

Now we consider the frequency spectrum of the soft fields.  When
$p_0\sim gT$, one can use perturbation theory to compute $A(P)$ and
$W(P)$ from their initial values. Consequently, $A(P)$ and $W(P)$ can
be estimated from the solution to the linearized kinetic equations
(see \kapitel{sec.solving.eom}). For $p_0\sim g^2 T$, perturbation
theory breaks down. Still, the solutions to the linearized kinetic
equations give the correct order of magnitude estimates. The results
are listed in Table \ref{tab.F.29}.

\begin{table}
\begin {center}
\tabcolsep 10pt
\renewcommand {\arraystretch}{1.95}
\begin{tabular}{|c|c|c|}
\hline
$p_0$  &    $A(P)$      &  $W(P,\vv)$  
\\ \hline\hline
$g^4T$ & $g^{-9}T^{-3}$ & $g^{-7}T^{-3}$ 
\\ \hline
$g^2T$ & $g^{-6}T^{-3}$ & $g^{-6}T^{-3}$ 
\\ \hline                                
$g T $ & $g^{-5}T^{-3}$ & $g^{-5}T^{-3}$
\\
\hline
\end{tabular}
\vspace{.7cm}
\end {center}
\caption{Order of magnitude estimates for the amplitudes of the
Laplace transformed soft fields.}
\vspace{.7cm}
\mlabel{tab.F.29}
\end{table}

Inspecting Table \ref{tab.F.29}, one can see that the difference
$\delta A(t,\vp)=A(t,\vp)-A(0,\vp)$, which can be estimated as $p_0
A(P)$, is small compared with $A(0,\vp)$, as long as $p_0\gsim g^2
T$. In order to obtain large values of $\delta A(t,\vp)$, one has to
consider smaller frequencies.  In \rf{asy} the frequency scale
associated with large $\delta A(t,\vp)\sim A(0,\vp)$ was estimated as
\begin{eqnarray}
        \mlabel{g4t}
        p_0 \sim g^4 T
	,
\end{eqnarray}
again by considering the linearized kinetic equations. Later on, we
will see that there is a logarithmic correction to this estimate. In
this Section, we are only concerned about powers of the coupling and
we can use \mref{g4t}. The gauge fields are saturated at these small
frequencies. Therefore we have
\begin{eqnarray}
  \mlabel{estimate.A}
  \vA(P)\sim  g^{-9} T^{-3}  
	\qquad (p_0 \sim g^4 T)
	.
\end{eqnarray}
The small frequency part of $W$ cannot be estimated from the
linearized kinetic equations. Therefore, one has to jump ahead and use
the result from the effective theory for the soft field modes (see
\kapitel{sec.log.scales}). One finds
\begin{eqnarray}
        W(P,v)\ \sim  g^{-7} T^{-3}
	\qquad (p_0 \sim g^4 T)
	.
\end{eqnarray}
Note that this is smaller than $A(P)$ by two powers of the gauge
coupling. From Table \ref{tab.F.29}, one can see that the frequency
spectra of $A$ and $W$ are qualitatively different. The main
contribution to $A(t)\sim \int d p_0 A(p_0)$ comes from $p_0\sim g^4
T$, while $W(t)$ receives its main contributions from $p_0$ of order
$gT$ and $g^2 T$.

\msection{The perturbative expansion of $\xi$} \mlabel{sec.xiexpansion}
Compared with the scalar model discussed in \kapitel{sec.factorization},
the gauge theory case  is complicated by the following
circumstances.
\begin{enumerate}

\item There are several soft fields that can play the role of $\Phi$
(see \eq{F.1.1.5}).

\item The various soft fields have different characteristic
frequencies. The gauge field is ``slow'' in the sense that $A(t)$ is
saturated by frequencies of order $g^4 T$. In contrast, $W(t)$ is
``fast'' since $W(t)$ is mainly determined by frequencies of order
$gT$ and $g^2 T$. If the equations of motion were linear, this would
not be a problem. Then, only the low frequency tail of $W$ would be
relevant to the dynamics of the gauge fields. However, due to the
non-linearities, the high frequency parts of $W$ might sneak into the
low frequency parts of the equations of motion.

\item For the term $\xi_0$ the replacement \mref{replacechi} gives
zero due to the anti-symmetry of the structure constants of the gauge
group. Therefore, one has to keep the term $\xi_0$ in the equations of
motion for the soft fields. Due to the factorization property
\mref{chi.approximation}, correlators of $\xi_0$ can be approximated
by 2-point functions. In other words, $\xi_0$ acts on the soft fields
as a Gaussian noise.

\end{enumerate} 
\subsection{The term $\xi_0$} \mlabel{sec.xi0}
We are now ready to discuss the first term in the expansion
\mref{xiexpansion}, 
\begin{eqnarray}
        \mlabel{F.0.5.1.7}
        \xi^a_0 (x,\vv) = -g f^{abc} \( v\cdot a_0^b(x)
        w_0^c(x,\vv) \)_{\rm soft}
	.
\end{eqnarray}
In momentum space, it reads (cf.\ \eq{laplace.product})
\begin{eqnarray}
        \mlabel{xi0.general}
        \xi_0^a(P,\vv) = -g f^{abc} 
        \intsub{    0<\im k^0<\im p^0 }  
        \frac{d^4 k}{(2\pi)^4} v\cdot a_0^b(K) w_0^c(P - K,\vv)
	.
\end{eqnarray}
In \kapitel{sec.factorization} we have argued that one can
approximate the product $a_0 w_0$ by its thermal expectation
value. However, this replacement gives zero for $\xi_0$, because $
\langle a_0^b w_0^c \rangle $ is proportional to $\delta^{bc}$ which
is contracted with the anti-symmetric structure constant $f^{abc}$.
Therefore, we have to leave $\xi_0$ in Eq.~\mref{vlasovsoft} as it
stands and the solution to the equations of motion for the soft fields
will contain products of $\xi_0$'s. However, the thermal averages of
these products can be approximated by a product of 2-point functions,
\begin{eqnarray}
  \lefteqn{
  \langle \xi_0^{a_1}(x_1,\vv_1)\cdots
  \xi_0^{a_{n}}(x_{n},\vv_{n}) \rangle  \simeq }
\nn \\&&
  \langle \xi_0^{a_1}(x_1,\vv_1)
  \xi_0^{a_{2}}(x_{2},\vv_{2}) \rangle
  \cdots
  \langle \xi_0^{a_{n-1}}(x_{n-1},\vv_{n-1})
   \xi_0^{a_{n}}(x_{n},\vv_{n})\rangle
\nn \\
   \mlabel{xi0.n.point}
        &&
   {}+ \mbox{permutations}(1,\ldots,n)
	,
\end{eqnarray}
which means that $\xi_0$ acts as a Gaussian noise.

For the following it will be important to know the characteristic amplitude of
$\xi_0$. Since $\langle \xi_0\rangle$ vanishes and because the
leading order contribution is determined by its 2-point function, the
amplitude of $\xi_0$ is determined by
\begin{eqnarray}
        \xi_0 \sim \langle\xi_0 \xi_0 \rangle^\half
	.
\end{eqnarray}
Inspecting \eq{F.0.5.1.7} one sees that $\xi_0(P)$ is of the form
\begin{eqnarray}
        \mlabel{F.31.1}
        \xi_0 (P)\sim g \int d^4 k \chi \sim g (gT)^4 \chi
	,
\end{eqnarray}
where $\chi$ is a product of semi-hard fields analogous to \mref{chi}.
Therefore, one can apply the arguments of \kapitel{sec.factorization},
\begin{eqnarray}
        \mlabel{F.31.2} 
	\langle\xi_0(P) \xi_0(P') \rangle \sim 
	g^2
        (gT)^8 \langle\chi \chi \rangle_{\rm conn.} 
	\sim g^2
        (gT)^8\(g^2 \frac{p_0}{T} \) \langle \chi\rangle^2
	,
\end{eqnarray}
where we have used \eq{estimatechichi.2} in the second step. The 
characteristic amplitude of $\langle \chi\rangle$ was determined in 
\eq{estimatechi}. Inserting the result gives
\begin{eqnarray}
        \mlabel{F.31.3}
        \langle\xi_0(P) \xi_0(P') \rangle \sim 
        p_0^{-1} g^{-6} T^{-3}
	,
\end{eqnarray}
so that we finally obtain 
\begin{eqnarray}
        \mlabel{F.31.4}
        \xi_0 (P) \sim 
        \(p_0^{-1} g^{-6} T^{-3}\)^\half
	.
\end{eqnarray}
For the characteristic frequency of the soft non-perturbative dynamics
\mref{g4t} this gives
\begin{eqnarray}
        \mlabel{F.31.5} 
	\xi_0(P) \sim g^{-5} T^{-2} 
	\qquad (p_0\sim g^4 T)
	 .
\end{eqnarray}

We will now determine the expansion parameter which controls the
accuracy of the Gaussian approximation \mref{xi0.n.point} for
$\xi_0$. By using \mref{xi0.n.point} one neglects the contribution due
to the 3-point function
\begin{eqnarray}
        \[\xi_0\]_{\mbox{\scriptsize 3-point function}} \sim 
        \langle \xi_0\xi_0\xi_0 \rangle^\frac13
	.
\end{eqnarray}
For the rhs can one can use the arguments of \kapitel{sec.factorization}, 
\begin{eqnarray}
        \mlabel{F.32.1}
        \langle \xi_0(P)\xi_0(P')\xi_0(P'') \rangle 
        &\sim& 
        g^3 (gT)^{12}   \langle\chi \chi \chi   \rangle_{\rm conn.}
\nn\\
        &\sim&
        g^3 (gT)^{12} \(g^2 \frac{p_0}{T}\)^2 \langle \chi \rangle^3
	.
\end{eqnarray}
Comparing this to \eq{F.31.4}, we find 
\begin{eqnarray}
        \mlabel{F.32.4}
        \[\xi_0(P)\]_{\mbox{\scriptsize 3-point function}}         
        \sim
        \(g^2 \frac{p_0}{T}\)^\frac16 \xi_0(P)
	.
\end{eqnarray}
For the frequency \mref{g4t} this gives
\begin{eqnarray}
        \mlabel{F.32.5}
        \[\xi_0(P)\]_{\mbox{\scriptsize 3-point function}}         
        \sim
        g \xi_0(P)
	\qquad (p_0\sim g^4 T)
	.
\end{eqnarray}
Thus, we find that the Gaussian approximation \mref{xi0.n.point} for 
$\xi_0$ neglects terms which are suppressed by one power of $g$.
\subsection{The term $\xi_1$} \mlabel{sec.xi1}
Now consider the second term in the expansion \mref{xiexpansion} of
$\xi$,
\begin{eqnarray}
	\mlabel{xi1}
        \xi_1 (x,\vv) = i g \Big\{
        [v\mal a_1(x), w_0(x,\vv)] + [v\mal a_0(x), w_1(x,\vv)]
        \Big\}_{\rm soft}
	.
\end{eqnarray}
The parametric form of $\lav\xi_1\rav$ was determined in
\kapitel{sec.factorization}, the result was \eq{F.19.3}. Now we have
two soft fields, $W$ and $A$, which can play the role of $\Phi$.  A
non-zero contribution due to $A$ would spoil the gauge covariance of
the equation of motion for the soft fields \mref{vlasovsoft}. However,
as we have argued in \kapitel{sec.separation}, $\xi$ transforms
covariantly under soft gauge transformations and such a non-covariant
term cannot occur (for an illustration of how these contributions
drop out, see Appendix C).  Thus, at leading
order in $g$ we have
\begin{eqnarray}
        \lav\xi_1\rav \sim g^2 T W
	.
\end{eqnarray}
Taking into account the dependence on $\vv$, the most general form
linear in $W$ is
\begin{eqnarray}
        \mlabel{xi1.general}
        \lav\xi_1(x,\vv)\rav = g^2 T  \intv{1} {\cal I}
        (\vv,\vv_1) W(x,\vv_1) 
	.
\end{eqnarray}
The amplitude of $\lav\xi_1\rav$ for the frequency scale $p_0 \sim g^4
T$ is of order (cf.\ Tab.~\ref{tab.F.29})
\begin{eqnarray}
        \mlabel{size.of.xi1}
        \lav\xi_1(P,\vv)\rav \sim g^{-5} T^{-2}
	.
\end{eqnarray}
Thus, we find that $\xi_0(P,\vv)$ and $\lav\xi_1(P,\vv)\rav$ are of the
same order of magnitude when $p_0$ is the characteristic frequency of
the soft gauge fields \mref{g4t}.

In \kapitel{sec.factorization} it was argued that the leading order
contribution to $\xi_1$ is given by 
\begin{eqnarray}
	\mlabel{xi1.app}
	\xi_1 \simeq \lav\xi_1\rav
	,
\end{eqnarray}
while higher correlation functions can be neglected. Here the
situation is slightly more complicated because we have several soft
fields and these have different characteristic amplitudes. In particular,
the leading order contribution to $\lav\xi_1\rav$ found above is
relatively small because $W$ is smaller than $A$. In addition, the
soft fields have different characteristic frequencies. Thus, the
frequency in the estimate \mref{estimate.delta} may be different from
$p_0\sim g^4 T$. 

Still, one can use the arguments of \kapitel{sec.factorization} to
show that corrections to \mref{xi1.app} are negligible. Since $A(P)$
is larger than $W(P)$ for any $p_0$ (see Tab.~\ref{tab.F.29}), one can
simply ignore $W$ and apply the arguments of
\kapitel{sec.factorization} to the field $A$ only. This simplifies the
discussion because $A$ is dominated by small frequencies of order
$g^4 T$. Imagine that $A$ gives a non-vanishing contribution to $\lav
\xi_1 \rav $. This would be of order
\begin{eqnarray}
      \lav \xi_1 (P,\vv) \rav_A \sim  
       g^{-7} T^{-2}
	.
\end{eqnarray}
The largest possible contribution (gauge covariant or not) to the
connected 2-point function of $\xi_1$ can then be estimated using
\mref{estimatechichi.2},
\begin{eqnarray}
      \lav \xi_1 \xi_1 \rav_{A} \sim  
       \(g^2\frac{p_0}{T}\)\( \lav \xi_1 \rav_{A} \)^2  
       \sim g^{-8} T^{-4}
	,
\end{eqnarray}
which implies that 
\begin{eqnarray}
        \mlabel{F.46.5.5}
        \[\xi_1 (P) \]_{\mbox{\scriptsize 2-point function}}         
        \sim    g^{-4} T^{-2}
	.
\end{eqnarray}
Comparing this with \eq{size.of.xi1} we find
\begin{eqnarray}
        \mlabel{F.46.5.6}
        \[\xi_1 (P) \]_{\mbox{\scriptsize 2-point function}}         
        \sim    g \lav \xi_1 (P)\rav
	.
\end{eqnarray}
Thus, by using the approximation \mref{xi1.app} one neglects
contributions to $\xi_1$ which are suppressed by at least a factor
$g$.

\subsection{$\xi_n$ for $n\ge 2$}\mlabel{sec.higher.order}
In this Section it is shown that $\xi_n$ does not contribute to the
equations of motion for the soft fields at leading order in $g$ when
$n\ge 2$ . The estimates made in this Section are not meant to be
exact. Instead, they are to be understood as upper limits.

Let us first consider the term $\lav\xi_n\rav $.  The largest term
which is allowed by gauge covariance is
\begin{eqnarray}        
        \mlabel{F.48.1}
        \lav\xi_n\rav \sim g^n (gT)^{2 - n} W^n
	.
\end{eqnarray}
In momentum space it is of the form (ignoring the dependence on
$\vv$) 
\begin{eqnarray}        
        \lav\xi_n(P)\rav &\sim & g^n (gT)^{2 - n} 
        \int d^4p_1\cdots\int d^4p_{n-1} \nn
\\
\mlabel{F.48.2} &&
        W(P_1)\cdots W(P_{n-1}) W(P-P_1\cdots -P_{n-1})
	.
\end{eqnarray}
Note that not only the small frequency ($p_0\sim g^4T$) piece of $W$
contributes to this integral. We can estimate $d^4 p_i W(P_i)\sim W(x)
\sim g^2 T$. An upper limit for the last factor in \mref{F.48.2} is
$W(p_0\sim g^4 T,\vp) \sim g^{-7} T^{-3}$. Consequently,
\begin{eqnarray}        
        \mlabel{F.48.3}
        \lav\xi_n(P)\rav \sim g^{2n-7} T^{-2}
	,
\end{eqnarray}
which shows that $\lav\xi_n(P)\rav$ for $n\ge 2$ is in fact negligible
relative to $\xi_0$ and $\lav\xi_1\rav$.

Estimates for connected correlation functions of $\xi_n$ can be
obtained in the same way as for $\xi_1$ in \kapitel{sec.xi1}. First,
we estimate the amplitude of the would-be contribution of $A$ to
$\lav\xi_n\rav$. Proceeding as for $W$, we find
\begin{eqnarray}        
        \mlabel{F.48.2.2}
        \lav\xi_n(P)\rav_A \sim g^{n-8} T^{-2}
	.
\end{eqnarray}
The connected 2-point function of $\xi_n$ can now be estimated using
\eq{estimatechichi.2}. For $p_0\sim g^4 T$ one finds
\begin{eqnarray}        
\mlabel{F.48.2.1}
        \[\xi_n (P) \]_{\mbox{\scriptsize 2-point function}}         
        \sim  \(g^2 \frac{p_0}{T}\) ^{\frac12}
        \lav\xi_n(P)\rav_A 
        \sim g^3    \lav\xi_n(P)\rav_A 
	.
\end{eqnarray}
Consequently, 
\begin{eqnarray}
        \mlabel{F.48.2.3}
        \[\xi_n (P) \]_{\mbox{\scriptsize 2-point function}}         
        \lsim  g^{n-5} T^{-2}
	,
\end{eqnarray}
which is smaller than $\xi_0$ and $\xi_1$ by at least a factor $g^2$
when $n\ge 2$.
  
\subsection{The Boltzmann equation for the soft fields}  
\mlabel{sec.boltzmann}
To summarize this Section, we have found that at leading order in the
gauge coupling, the equation of motion for the soft fields
\mref{vlasovsoft} can be approximated as 
\begin{eqnarray}
        \mlabel{vlasovsoft.3}
        [v \cdot D, W(x,\vv)] \simeq \vec{v}\cdot\vec{E} (x)
        + \xi_0(x,\vv) + \lav\xi_1(x,\vv)\rav
	.
\end{eqnarray}
Here $\xi_0$ is a Gaussian noise due to \eq{xi0.n.point}.  Since the
expectation value of $\xi_0$ vanishes, we also had to keep the second
term of the expansion \mref{xiexpansion}, which can be approximated as
in \mref{replacexi}.  The terms $\xi_n$ with $n\ge 2$ can be
neglected. The form of $ \lav\xi_1\rav$ is given by
\mref{xi1.general}, so that \eq{vlasovsoft.3} can be written as
\begin{eqnarray}
        \mlabel{vlasovsoft.4}
        [v \cdot D, W(x,\vv)] &\simeq& \vec{v}\cdot\vec{E} (x)
	+ \xi_0(x,\vv)
\nn\\ && 
        {} + g^2 T  \intv{1} {\cal I}
        (\vv,\vv_1) W(x,\vv_1) 
	.
\end{eqnarray}
This equation has the form of a linearized Boltzmann equation for $W$,
which contains a Gaussian noise term. The third term on the rhs plays
the role of a collision term (for a detailed discussion of this
equations see \kapitel{sec.log.boltzmann}).

\msection{The leading log approximation} \mlabel{sec.logarithmic}
The noise and the collision term in \mref{vlasovsoft.4} contain
contributions which are logarithmically sensitive to the separation
scale $\mu$ in \eq{mu}. These contributions will be calculated in this
Section. They are due to the transverse semi-hard gauge
fields. Therefore, one can use the expressions for $\va_0$ and $w_0$
which were obtained in \kapitel{sec.solving.eom} and ignore the
longitudinal fields. Then, it will be argued that, in order to obtain
the effective theory for the soft field modes at leading logarithmic
order, the logarithm can be replaced by $\log(1/g)$. In this
approximation the Boltzmann equation can be solved. This determines
the current on the rhs of the Maxwell equation \mref{maxwellsoft}. For
the non-perturbative dynamics of the soft modes, the kinetic term in
\mref{maxwellsoft} can be neglected which then leads to the Langevin
equation \mref{langevintag}.

\subsection{The 2-point function of $\xi_0$}
To compute the 2-point function of $\xi_0$, one can use
\eq{xi0.general} and insert the results for the correlators of $\va_0$
and $w_0$ which are listed in \eqs{aa}-\mref{ww}. It is simpler,
however, to take advantage of translation invariance in time since
$\xi_0$ does not depend on the soft background.  Then one can proceed
as in \eqs{F.13.1}, \mref{F.13.2} and write
\begin{eqnarray}
  \mlabel{xi0correlator1}
    \left\langle 
    \xi_0^{a}(P_1,\vv_1)
    \xi_0^{b}(P_{2},\vv_{2}) 
    \right\rangle 
    &=&
    \frac{i}{p_1^0 + p_2^0}
    \[
    \left\langle 
    \xi_0^{a}(P_1,\vv_1)
    \xi_{\rm in}^{b}(\vp_{2},\vv_{2}) 
    \right\rangle   
    +
    (1\leftrightarrow 2)
    \]
	,
\end{eqnarray}
where the subscript ``in'' denotes the initial value at $t=0$. To
evaluate the rhs of \mref{xi0correlator1} one needs the correlations
of $\va_0$ and $w_0$ with the corresponding initial values; these
are listed in \eqs{aain}-\mref{awin}. One obtains\footnote{In this
Section, ``$\approx$'' indicates that these equation are valid with
logarithmic accuracy.}
\begin{eqnarray}
  \mlabel{xi.2.1}
    \left\langle 
    \xi_0^{a}(P_1,\vv_1)
    \xi_{\rm in}^{b}(\vp_{2},\vv_{2}) 
    \right\rangle &\approx& - \frac{Ng^2 T^2}{\mmdebye}\, J(P_1,\vv_1,\vv_2)
	\delta^{ab} (2\pi)^3 \delta^3 (\vp_1 + \vp_2) 
	,
\end{eqnarray}
where $J$ is the integral
\begin{eqnarray}
  \mlabel{integral}
  J (P,\vv,\vv') &=& v^i v_1^j 
        \!\!
        \intsub{0<\im k^0<\im p^0 }
        \!\!
        \frac{d^4 k}{(2\pi)^4}
  \left\{ \frac{1}{\mmdebye} 
  \Delta_{12}^i(K,\vv') \Delta_{12}^j(P- K,\vv) \right. 
\\
  && 
	{} + \left. \nn
  \frac{i}{p^0 - k^0} 
  \( \Delta_{11}^{ij}(P- K) - \Delta_{11}^{ij}(0,\vp - \vk) \)
  \Delta_{22}(K,\vv,\vv')
        \right\}
	\nn
	.
\end{eqnarray}
After neglecting the soft momentum $P$, this integral is
logarithmically divergent for $k\to 0$ (see below).  Since we are
integrating out only fields with $k>\mu$, we will find a contribution
proportional to $\log(1/\mu)$.  The integral is convergent in the
ultraviolet, so that the only other scale in \mref{integral} is
$\mdebye\sim gT$.  Therefore, the $\mu$ dependent part of
\mref{integral} will be proportional to $\log(gT/\mu)$.  Here, we will
only compute this logarithmic contribution to (9.3).

To see how one obtains the logarithmic contribution to
\mref{integral}, consider the transverse propagator
\mref{propagator}. For $|k^0|\sim k$, the transverse self-energy
\mref{polarization} is of order  $\mmdebye$ and there is no infrared
divergence. Therefore, we must consider $|k^0|\ll k$. In this limit, 
the self-energy \mref{polarization} can be approximated by 
\begin{eqnarray}
        \mlabel{deltapilimit} 
        \delta \Pi_\tr (K) \simeq 
	-i \frac{\pi}{4} \mmdebye \frac{k_0}{k} \qquad (|k_0|\ll k)
	.
\end{eqnarray}
Then, the hard thermal loop resummed propagator \mref{propagator}
becomes
\begin{eqnarray}
      \mlabel{int.1.2}
        \stern\Delta_{\rm t}(K) \simeq \frac{1}{k^2}
        \frac{i \gamma_\vk}{k^0 + i \gamma_\vk}
           \qquad (|k^0| \ll k)
	,
\end{eqnarray}
where
\begin{eqnarray}
        \gamma_\vk = \frac{4}{\pi} \frac{k^3}{\mmdebye}
	.
\end{eqnarray}
Thus, for $|k^0| \lsim \gamma_\vk$ the transverse propagator is of
order $1/k^2$, which means that it is unscreened.  The infrared
divergence occurs only in the transverse sector. The longitudinal hard
thermal loop self-energy stays of order $\mmdebye$ when $k_0\to 0$.

With this approximation, assuming $|p^0| \lsi g^2 T$, and neglecting
$\vp$ in $\Delta^{ij}_{11}(K - P)$ the first term in
Eq.~\mref{integral} becomes
\begin{eqnarray}
        \mlabel{integral1}
        &&
          v^i v'^j 
        \!\!
        \intsub{0<\im k^0<\im p^0 }
        \!\!
        \frac{d^4 k}{(2\pi)^4}
  \frac{1}{\mmdebye} 
  \Delta_{12}^i(K,\vv') \Delta_{12}^j(P- K,\vv) \approx   
\\
         &&\mmdebye 
        \!\!
        \intsub{0<\im k^0<\im p^0 }
        \!\!
        \frac{d^4 k}{(2\pi)^4}
        \frac{\gamma_\vk^2}{k^4} 
        \(\vv_\tr\mal\vv'_\tr\)^2 
        \frac{1}{v'\cdot K} \frac{1}{k^0 + i \gamma_\vk}
        \frac{1}{v\cdot (K - P)} \frac{1}{k^0 - p^0 - i \gamma_\vk}
        \nn
	.
\end{eqnarray}
The $k^0$ integration contour can be closed in the lower half plane
picking up the poles at $k^0 = \vv_1\cdot\vk$ and $k^0 = - i
\gamma_\vk$. Note that this is consistent only as long as
$k^2\ll\mmdebye$. Only then  we have $\gamma_\vk\ll k$, which is
required for $|k^0| \ll k$.  Then, we can neglect $p^0$ and $\vp$ in
Eq.~\mref{integral1} except for the imaginary part of $p^0$. We can
thus set $p^0 \to i\epsilon$. Proceeding similarly for 
the second term in Eq.~\mref{integral}, we find
\begin{eqnarray}
        \lefteqn{
        J (P,\vv,\vv') \approx i \intkd{} \left\{
         \frac{1}{k^2}
        \, \frac{1}{\vv\cdot\vk - i\gamma_\vk} 
        \, \vv_\tr^2 
        \, 4\pi \deltav(\vv - \vv')  \right.
         }
\nn\\ 
	\mlabel{integral2}
        && 
	{}- \left. \mmdebye \frac{\gamma_\vk^2}{k^4}
        \(\vv_\tr\mal\vv'_\tr\)^2
        \frac{1}{\vv'\cdot\vk + i\gamma_\vk}
        \, \frac{1}{\vv'\cdot\vk - i\gamma_\vk}
        \, \frac{1}{\vv'\cdot\vk - \vv\cdot\vk -i\epsilon }
        \right\}
	.
\end{eqnarray}
The first integral in \mref{integral2} is
\begin{eqnarray}
        \intkd{} 
         \frac{1}{k^2}
        \, \frac{1}{\vv\cdot\vk - i\gamma_\vk} 
        \, \vv_\tr^2 
        &=&
        \frac{1}{(2\pi)^2}
        \int_\mu^{gT} d k \int_{-1}^{1} d y\frac{1-y^2}{yk- i\gamma_\vk}
        \nn \\
        \mlabel{int.4.2}
        &\approx &\frac{i}{4\pi} \log\(\frac{gT}{\mu}\) 
	.
\end{eqnarray}
Here, the upper limit for the $k$ integration was chosen as $ gT$
which is where the low frequency approximation \mref{int.1.2} ceases
to be valid.  For the second integral we choose $\vv_1$ as the
3-axis. The 1-axis is chosen such that $\vv$ lies in the 1-3 plane. As
in \eq{int.4.2}, the integral over the polar angle is saturated at
small $y=\cos(\theta)$. Then, we have $\vv_\tr\mal\vv_{1, \tr} \simeq
\vv\mal\vv_1$. For the  integral over the azimuthal angle $\varphi$ of 
$\vk$ one obtains
\begin{eqnarray}
	\mlabel{yintegral}
      \int_{-\pi}^\pi \frac{d\varphi}{\vv'\cdot\vk - \vv\cdot\vk -i\epsilon }
      \approx \frac{1}{k} \frac{2\pi i}{\sqrt{1 - (\vv\mal\vv')^2}}
	.
\end{eqnarray}
The remaining integral over the polar angle gives
\begin{eqnarray}
	\mlabel{phiintegral}
      \int_{-1}^1 d y          
      \frac{1}{y k + i\gamma_\vk}
        \, \frac{1}{y k - i\gamma_\vk}
        \approx \frac{\pi}{k   \gamma_\vk    }
	.
\end{eqnarray}
Both in \mref{yintegral} and \mref{phiintegral} the angular
integration is saturated at $\vv\mal\vk\sim\vv'\mal\vk\sim\gamma_k$,
which means that the approximation \mref{int.1.2} was consistent.
Again, the integral over $k$ gives a logarithm $\log(g T/\mu)$, so
that we finally obtain
\begin{eqnarray}
        J (P,\vv,\vv') \approx
        \log\(\frac{g T}{\mu}\)I(\vv,\vv')
	,
\end{eqnarray}
where
\begin{eqnarray}
  \mlabel{kern}
  I(\vv,\vv') = -\deltav(\vv - \vv') 
        + \frac{1}{\pi^2}
        \frac{(\vv\mal\vv')^2}{\sqrt{1 - (\vv\mal\vv')^2}} 
	.
\end{eqnarray}
The 2-point function of $\xi_0$ can therefore be written as
\begin{eqnarray}
  \mlabel{xi0correlator3}
    \left\langle 
    \xi_0^{a}(P,\vv)
    \xi_0^{b}(P',\vv') 
    \right\rangle 
    &\approx &
    -2 N \frac{g^2 T^2}{\mmdebye}
        \log\(\frac{g T}{\mu}\)I(\vv,\vv')
     \nn\\ &&
     \delta^{ab}     \frac{i}{p_0 + p'_0}
     (2\pi)^3 \delta^3 (\vp + \vp')
	.
\end{eqnarray}
In configuration space, this corresponds to
\begin{eqnarray}
        \mlabel{xi0correlator4}
        \left\langle 
    \xi_0^{a}(x,\vv)
    \xi_0^{b}(x',\vv') 
    \right\rangle 
    \approx
    -2  \frac{Ng^2 T^2}{\mmdebye}
        \log\(\frac{g T}{\mu}\)
         I(\vv,\vv')
        \delta^{ab} \delta^4(x - x')
	.
\end{eqnarray}

\subsection{The collision term $\lav\xi_1\rav$} \mlabel{sec.xi1.log}
In this Section we approximate $\xi_1$ according to Eq.~\mref{replacexi},
i.e.\ we replace the products of the free semi-hard fields $\va_0$ and
$w_0$ by their expectation values. We only need to take
into account the piece of $\xi_1$ which contains the field $W$. The 
contributions due to $A$ vanish after the replacement~\mref{replacexi}
(see  \kapitel{sec.xi1} and Appendix C). Then we have 
\begin{eqnarray}
        \xi_1^a(P,\vv) &\approx& g^2 f^{abc} 
        \!\!
        \intsub{0<\im k^0<\im p^0 }
        \!\!
        \frac{d^4 k}{(2\pi)^4}
        \intv{1} 
        \!\!
        \intsub{0<\im p_1^0<\im k^0 }
        \!\!
        \frac{d^4 p_1}{(2\pi)^4} W^e(P_1,\vv_1)  
        \nn\\ 
        &&  \[ 
        f^{bde} v^i \Delta_{12}^{i}(K,\vv_1) 
        \vv_1\cdot \va_0^d(K-P_1)  w_0^c(P - K,\vv)       \right. \nn\\
        && 
        \left.
        {}+  f^{cde}  \Delta_{22}(K,\vv,\vv_1) 
         \vv\cdot\va_0^b(P-K)   \vv_1\cdot\va_0^d(K-P_1)  \] 
	.
\end{eqnarray}
Using the results for the 2-point functions of $\va_0$ and $w_0$ in
\mref{aa} and \mref{aw} we find
\begin{eqnarray}
\lefteqn{
  \lav\xi_1^a(P,\vv)\rav \approx N g^2 T  
        \!\!
        \intsub{0<\im k^0<\im p^0 }
        \!\!
        \frac{d^4 k}{(2\pi)^4}
        \intv{1} 
        \!\!
        \intsub{0<\im p_1^0<\im k^0 }
        \!\!
        \frac{d^4 p_1}{(2\pi)^4}
        \frac{i}{p^0 - p_1^0}
        (2\pi)^3 \delta^3 (\vp - \vp_1)
         }\nn\\ 
        && W^a(P_1,\vv_1) \Bigg\{     
        v^i \Delta_{12}^{i}(K,\vv_1) v_1^j \frac{1}{\mmdebye}
        \[\Delta_{12}^{j}(K-P_1,\vv) + \Delta_{12}^{j}(P-K,\vv) \]
        \nn\\ && 
        \phantom{\mbox{$ppppp$}} 
        {} + \Delta_{22}(K,\vv,\vv_1)v^i v_1^j 
        \[ \frac{i}{p^0-k^0}
        \( \Delta_{11}^{ij}(P-K) - \Delta_{11}^{ij}(0,\vp-\vk)\)
        \right.
        \nn\\ 
  \mlabel{xi1replace1}
        && {} 
        \phantom{\mbox{$\Delta_{22}(K,\vv,\vv_1)v^i v_1^j$}} 
        \left.
         + \frac{i}{k^0-p_1^0}
        \( \Delta_{11}^{ij}(K-P_1) - \Delta_{11}^{ij}(0,\vk - \vp_1)\) \]
\Bigg\}
	.
\end{eqnarray}
The terms in the curly bracket which do not depend on $P$ vanish
because for them the $k^0$-integration contour can be moved to
$+i\infty$ without hitting any cut or pole. The remaining terms in the
curly bracket do not depend on $P_1$. Therefore, the $p_1^0$-contour
can be closed in the upper half plane which gives a contribution due
to the pole at $p_1^0 = p^0$,
\begin{eqnarray}
  \mlabel{xi.7.1}
   \!\!\!\!\!\!\!\!\!\!
        \int\limits_{
          {}^{ \phantom{D}}
          _{\scriptstyle 0<\im p_1^0<\im p^0} }
          \!\!\!\!\!\!\!\!\!\!\!
        \frac{d p_1^0}{2\pi}  \frac{i}{p^0 - p_1^0}W(p_1^0,\vp,\vv_1) 
        = W(P,\vv_1) 
	.
\end{eqnarray}
Then we obtain
\begin{eqnarray}
  \mlabel{xi1replace2}
   \lav \xi_1(P,\vv) \rav \approx N g^2 T   \intv{1} J(P,\vv,\vv_1) W(P,\vv_1)
	,
\end{eqnarray}
where $J(P,\vv,\vv_1)$ is precisely the integral \mref{integral} we
encountered in the previous Subsection. With the approximations discussed
there, we finally find
\begin{eqnarray}
  \mlabel{xi1replace3}
   \lav\xi_1(P,\vv)\rav \approx N g^2 T  \log\(\frac{gT}{\mu}\)
 \intv{1} I(\vv,\vv_1)        
    W(P,\vv_1) 
	.
\end{eqnarray}

\subsection{The Boltzmann equation}\mlabel{sec.log.boltzmann}
With the results obtained above we can write \eq{vlasovsoft.4} as
\begin{eqnarray}
        \mlabel{boltzmann}
        [v \cdot D, W(x,\vv)] &\approx& \vec{v}\cdot\vec{E} (x)
        + \xi_0(x,\vv) \nn\\ &&
        {} + N g^2 T  \log\(\frac{g T}{\mu}\)
        \intv{1}   I(\vv,\vv_1)      
        W(x,\vv_1)
	.
\end{eqnarray}
Eq.~\mref{boltzmann} is a Boltzmann equation for the soft fluctuations
of the particle distribution $W$. The term $\xi_0$ is independent of
the fields and therefore acts as a stochastic force. The third term on
the rhs of Eq.~\mref{boltzmann} has the form of a collision term. It
is due to the interactions with the semi-hard fields.

Let us see whether Eq.~\mref{boltzmann} is consistent with the Maxwell
equation \mref{maxwellsoft} for the soft fields. This requires that the
current
\begin{eqnarray}
        \mlabel{current}
   J^\nu(x) = \mmdebye \int\frac{d\Omega_\vec{v}}{4\pi}
        v^\nu  W(x,\vv)
\end{eqnarray}
is covariantly conserved.  Integrating Eq.~\mref{boltzmann} over the
direction of $\vv$, the third term on the rhs drops out due to
\begin{eqnarray}
        \mlabel{vanish}
        \intv{} I(\vv,\vv_1) = 0 
	,
\end{eqnarray}
and one obtains
\begin{eqnarray}
\mlabel{nonconservation}
        D_\mu J^\mu(x) = \mmdebye \intv{} \xi_0(x,\vv) 
	.
\end{eqnarray}
That is, the current appears not to be conserved. However,  we have
argued that only the 2-point function of $\xi_0$ should be relevant to the
leading order behaviour of the soft gauge fields. But the 2-point
function of $\xi_0(x_1,\vv_1)$ with the rhs of \eq{nonconservation}
vanishes,
\begin{eqnarray}
        \left\langle \xi_0(x_1,\vv_1) \intv{} \xi_0(x,\vv) \right\rangle =0,
\end{eqnarray}
due to \eq{vanish}. Thus, we can replace
\begin{eqnarray}
        \mlabel{intxi0}
        \intv{} \xi_0(x,\vv) \simeq 0
	,
\end{eqnarray}
so that the current is covariantly conserved within the present
approximation.

Due to the term $\xi_0$, Eq.~\mref{boltzmann} looks as if it was not
gauge covariant. However, the 2-point function of $\xi_0$,
\eq{xi0correlator4}, is invariant under gauge transformations of
$\xi_0$. Therefore, a gauge invariant correlation function computed
using \eq{boltzmann} will not depend on the choice of a gauge after
the average over $\xi_0$ has been performed. 

The only spatial momentum scales which are left in the problem are
$\mu$ and $g^2 T$. The field modes we are ultimately interested in
are the ones which have only momenta of order $g^2T$.  The cutoff
dependence on the rhs must drop out after solving the equations of
motion for the fields with spatial momenta smaller than $\mu$.  Thus,
after the $\mu$-dependence has cancelled, the logarithm in 
\mref{xi0correlator4} must turn into
$\log(gT/(g^2 T)) = \log(1/g)$. Therefore, at leading logarithmic
order the soft fields satisfy the Boltzmann equation
\begin{eqnarray}
        [v \cdot D,  W(x,\vv)] &=& \vec{v}\cdot\vec{E} (x)
        + \xi_0(x,\vv)  
\nn 
\\ 
        \mlabel{boltzmann.2}
        &&
        {}+ N g^2 T  \log\(1/g\)
         \intv{1}   I(\vv,\vv_1)      
         W(x,\vv_1)
	.
\end{eqnarray}
A collision term similar to the one in \mref{boltzmann.2}, has been
obtained previously by Selikhov and Gyulassy \cite{gyulassy}. It did
not contain the second term in \mref{kern} which, as we have seen
above, is necessary for the current \mref{current} to be
conserved. The collision term is larger by two powers of the coupling
than the collision term in an Abelian plasma.  This reflects the fact
that collisions in Abelian and non-Abelian plasmas are qualitatively
different, which was realized by Selikhov and Gyulassy
\cite{gyulassy}. The collision term in \mref{boltzmann.2} is dominated
by small angle scattering. In a non-Abelian plasma, even a single
small angle scattering event can change the color charge of the hard
particles, which is what is ``seen'' by the soft fields.\footnote{For
a recent discussion of earlier approaches to non-Abelian transport
equations, see~\rf{asy2}.}

\subsection{Solving the Boltzmann equation}
\mlabel{sec.solving.boltzmann} 
In the previous Subsection we have obtained a Boltzmann equation valid
for the soft field modes at leading order in $\log(1/g)$ when the
momenta in $W(x,\vv)$ are strictly of order $g^2 T$. It will now be
shown that one can neglect the lhs of Eq.\ \mref{boltzmann.2} up to
terms which are suppressed by a factor $(\log(1/g))^{-1}$. This is an
enormous simplification since it turns \mref{boltzmann.2} into a
relatively simple integral equation. It can then be solved for $W$,
which determines the current on the rhs of the Maxwell equation
\mref{maxwellsoft} for the soft gauge fields.

It is convenient to use a representation of the field $W(x,\vv)$ in
which the collision term is diagonal. Since the second term in
\mref{kern} depends only on $\vv_1\mal\vv_2$, the collision term
commutes with rotations of the velocity variable $\vv$. Thus, the
eigenfunctions of the collision term are the spherical harmonics
$Y_{lm}$ and its eigenvalues (see Appendix D) only depend on $l$.

The $l=1$ projection of the Boltzmann equation is
obtained by multiplying \eq{boltzmann.2} by $v^i$ and integrating
over $\Omega_\vv$. In $A_0=0$ gauge one obtains
\begin{eqnarray}
        \mlabel{boltzmann.wi}
        \partial_0 W^i(x) - \third [D^i, W(x)] - [D^j, W^{ij}(x)] &=& 
        \third E^i(x) + \xi_0^i(x) 
\\
        &&{}-  \frac{1}{4\pi}  N g^2 T\log\(1/g\) W^i(x)
	\nn
	.
\end{eqnarray}
Here,
\begin{eqnarray}
        \mlabel{w}
        W (x) &=& \intv{}  W(x,\vv) 
	,
\\
        \mlabel{wi}
        W^{i} (x) &=& \intv{} v^{i} W(x,\vv) 
	,
\\
        \mlabel{wij}
        W^{ij} (x) &=& 
        \intv{} \(v^{i}v^{j} -\third \delta^{ij}\) W(x,\vv) 
	,
\end{eqnarray}
contain the $l=0$, 1, 2 components of $W(x,\vv)$,
respectively. Similarly, 
\begin{eqnarray}
        \mlabel{xi0i}
        \xi_0^{i} (x) = \intv{} v^{i} \xi_0(x,\vv) 
\end{eqnarray}
is the $l=1$ projection of $\xi_0(x,\vv)$.
 
It is easy to see that the first two terms on the lhs of
\eq{boltzmann.wi} can be neglected. The first term contains a time
derivative.  We are only interested in the low frequency part of
\mref{boltzmann.wi}, for which this term is very small. It should be
noted that the first term can be neglected only if one is interested
in the dynamics of the gauge fields, since $W(x,\vv)$ is mainly
determined by frequencies larger than $g^4 T$. Now consider the second term
in \eq{boltzmann.wi}. Using the Gau{\ss} law
\begin{eqnarray}
        \mlabel{gauss}
        [D_i, E^i(x)] = \mmdebye W(x)
	,
\end{eqnarray}
it can be estimated as
\begin{eqnarray}
        [D^i, W(x)] \sim \frac{1}{\mmdebye} (g^2 T)^2 E^i(x) \sim g^2 E^i(x)
	,
\end{eqnarray}
which is small compared with the term $E^i$ on the rhs.

In the following, it will become clear that the third term on the
lhs of \mref{boltzmann.wi} can be neglected as well. Therefore,
\mref{boltzmann.wi} can be trivially solved for $W^i$, and one obtains
\begin{eqnarray}
        \mlabel{wi.solution}
        W^{i} (x) 
        \approx
	 \frac{4\pi}{N g^2 T  \log(1/g)} \(\third E^i(x) + \xi_0^i(x) \)
	,
\end{eqnarray}
which is the result of \cite{letter}. It should be noted (and will be
explained below) that the argument which was used in \cite{letter} to
justify this approximation was not rigorous.\footnote{I
thank Peter Arnold, Dam Son and Larry Yaffe for clarifying discussions
about this point.}

Before discussing this point more carefully, let us first solve
\eq{boltzmann.2} for the higher $l$ projections of $W(x,\vv)$ which
is simpler.  For this purpose it is convenient to expand $W(x,\vv)$ and
$\xi_0(x,\vv)$ in spherical harmonics,
\begin{eqnarray}
        \mlabel{spherical.w}
        W(x,\vv) &=& \sum_{lm} W_{lm}(x) Y_{lm}(\vv)
	,
\\
        \mlabel{spherical.xi}
        \xi_0(x,\vv) &=& \sum_{lm} \xi_{0,lm}(x) Y_{lm}(\vv)
	.
\end{eqnarray}
Consider the $(l,m)$ projection of \eq{boltzmann} for $l\ge 2$,
\begin{eqnarray}
        \int d\Omega_\vv Y^*_{lm}(\vv) [v\mal D, W(x,\vv)] &=& 
        \xi_{0,lm}(x) 
\nn \\
	\mlabel{boltzmann.lm}
	&&{}+ \frac{\lambda_l}{4\pi}N g^2 T  \log(1/g) W_{lm}(x) 
	.
\end{eqnarray}
Here $\lambda_l$ are the eigenvalues of the integral kernel
\mref{kern}. For $l\ge 2$ they are all non-zero (see Appendix D).  The
lhs is of order $g^2 T W_{l\pm 1, m\pm 1} $, while the rhs is larger
by a factor $\log(1/g)$. Thus, to leading logarithmic accuracy the
solution to \mref{boltzmann.lm} is
\begin{eqnarray}
        \mlabel{lm.solution}
        W_{lm}(x) \approx -\frac{4\pi}{\lambda_l} \frac{1}{N g^2 T  \log(1/g)} 
        \xi_{0,lm}(x) \qquad (l\ge 2)
	.
\end{eqnarray}
\eq{boltzmann.lm} shows that all $W_{lm}$ are of the same order of
magnitude when $l\ge 2$.

Now we return to the discussion of \eq{boltzmann.wi}.  In
\cite{letter} it was argued that the term $\propto \log(1/g) W^i$ is
large compared with the lhs because it is logarithmically enhanced.
This argument is correct for the transverse projections of $W^i$, but
it is not correct for the longitudinal part.  In \cite{asy3}, the
longitudinal projection $W^i_{\rm L}$ was defined as the part of
$W^i$ which contributes to $[D^i,W^i]$. $W^i_{\rm L}$ enters the $l=0$
projection of \mref{boltzmann.2}, which is the equation for current
conservation discussed in \kapitel{sec.log.boltzmann}. In terms of the
fields in \mref{boltzmann.wi} it reads
\begin{eqnarray}
	\partial_0 W(x) - [D ^i, W^i_{\rm L}(x)] = 0
	.
\end{eqnarray}
This equation implies that $W^i_{\rm L}$ is very small. In particular,
the term proportional to $\log(1/g)W^i_{\rm L}$ in the longitudinal
projection of \mref{boltzmann.wi} is much smaller than the term
$[D^j,W^{ij}]_{\rm L}$ on the lhs, contrarily to the claim in
\cite{letter}.  However, due to \mref{lm.solution}, even the term
$[D^j,W^{ij}]$ is smaller than $\xi_0$ by a logarithm. This means that
one can in fact neglect the lhs of \eq{boltzmann.wi}, which is the
assertion made above and which leads to \eq{wi.solution}.  An
alternative derivation of \eq{wi.solution} was recently presented by
Arnold, Son, and Yaffe \cite{asy3}.

One may wonder what the longitudinal projection of \eq{boltzmann.wi}
means. Due to the smallness of $W^i_{\rm L}$, the leading terms in the
longitudinal projection of \mref{boltzmann.2} are $E^i_{\rm
L}$ and $\xi^i_{0\rm L}$.  Therefore, these two terms must cancel each
other which gives
\begin{eqnarray}
	\mlabel{elong}
	[D^i, E^i] \approx -3 [D^i, \xi_0^i]
	.
\end{eqnarray}
On first sight this result appears a little puzzling. The purpose of
this Section was to solve the Boltzmann equation which should
determine $W(x,\vv)$ as a function of $\vE$ and $\xi_0$. Now we have
obtained a condition for $\vE$ itself. The reason is that we have made
use of the Gau{\ss} law to argue that $W(x)$ can be neglected in
\eq{boltzmann.wi}. Therefore, \eqs{wi.solution}, \mref{lm.solution},
and \mref{elong} must be considered as an approximate solution to the
set of equations consisting of both the Boltzmann equation and of the
Gau{\ss} law.

\subsection{The Langevin equation}                \mlabel{sec.langevin}
Inserting the current $J^i = \mmdebye W^i$ into the Maxwell
equation \mref{maxwellsoft} one obtains
\begin{eqnarray}
        \mlabel{langevin1}
        \ddot{ A}^i +  [D_j,F^{j i}] \approx 
        - \gamma \dot{A}^i + \zeta^i
	,
\end{eqnarray}
where the damping coefficient (or color conductivity, see \cite{asy2})
is given by
\begin{eqnarray}
        \gamma = \frac{4\pi}{3 }\frac{\mmdebye}{Ng^2 T\log(1/g)}
	.
\end{eqnarray}
The Gaussian white noise $\zeta^i$ is proportional to $\xi_0^i$, 
\begin{eqnarray}
        \zeta^i =  \frac{4\pi\mmdebye}{Ng^2T\log(1/g)}
         \xi_0^i
	.
\end{eqnarray}
The 2-point function of $\zeta^i$ can be obtained from
\eqs{xi0correlator4} and \mref{Y.2.2.3},
\begin{eqnarray}
        \mlabel{langevin2}
        \left\langle 
    \zeta^{ia}(x_1)
    \zeta^{jb}(x_{2}) 
    \right\rangle 
        =
    2 T \gamma
        \delta^{ij } \delta^{ab}  \delta^4(x_1 - x_2)
	.
\end{eqnarray}

The equation of motion \mref{langevin1} describes an over-damped
system.  To see this, let us estimate the second term on the lhs. It
contains two covariant derivatives of the gauge fields. Each
derivative is of order $g^2 T$ . Thus, this term can be estimated
as\footnote{This estimate does not rely on perturbation theory. For
the soft modes both terms in the covariant derivative $\partial_i - g
A_i$ are of the same order because of $A_i \sim gT$ (see
\kapitel{sec.scales}).}
\begin{eqnarray}
	\mlabel{lhs}
	[D_j,F^{j i}] \sim \vD^2 \vA \sim g^4 T^2 \vA
	.
\end{eqnarray}
For the damping term on the rhs we have
\begin{eqnarray}
	\mlabel{rhs}
	\gamma \dot{\vA} \sim \frac{T}{\log(1/g)} \frac{ \vA}{t}
	,
\end{eqnarray}
where $t$ is the characteristic time scale associated with the soft
gauge fields.  Comparing \mref{lhs} and \mref{rhs}, we find
\begin{eqnarray}
	\mlabel{timescale}
  	t^{-1}\sim g^4 \log(1/g)T 
	.
\end{eqnarray}
Therefore, the second time derivative in \mref{langevin1} is
negligible and the dynamics of the soft modes is correctly described
by the Langevin equation
\begin{eqnarray}
        \mlabel{langevintag}
        [D_j,F^{j i}] = 
        -\gamma \dot{A}^i + \zeta^i  
	.
\end{eqnarray}
The approximation of dropping the term $\ddot{\vA}$ has an important
effect. The effective theory described by \eqs{langevin2} and
\mref{langevintag} is no longer sensitive to an UV cutoff
\cite{asy2}. In other words, this theory is UV finite (aside from
the ``vacuum energy'' which does not affect correlation functions like
\mref{c}). It is equivalent to the stochastic quantization
\cite{parisi,zj} of Yang-Mills theory in 3 Euclidean dimensions
(in contrast to the usual stochastic quantization, the time variable
in \mref{langevintag} is the physical time). Zinn-Justin and Zwanziger
\cite{zjz} have shown that both equal-time and different-time
correlation functions are renormalized by the same counter-terms as
the static theory, except for a possible renormalization of the
damping coefficient $\gamma$. But 3-dimensional Yang-Mills theory is
UV finite and even the coefficient $\gamma$ is not renormalized in 3+1
dimensions.

\begin{figure}[t]
 
 
\hspace{.8cm}
\epsfysize=6.5cm
\centerline{
        \hspace{-.3cm}
        \epsfysize=5cm\epsffile{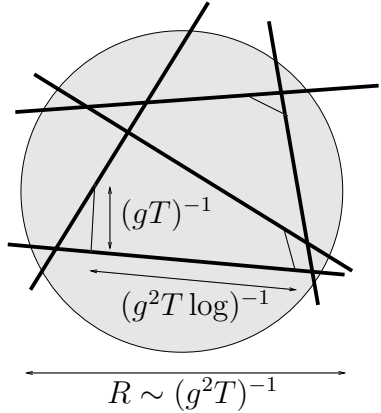}
        }

\vspace*{-3cm}
\begin{picture}(300,100)
%
\put(170,73){$(g T)^{-1}$}
\put(170,38){$(g^2 T\log)^{-1}$}
\put(165,5){$R\sim (g^2 T)^{-1}$}
\end{picture}

\caption[a]{The soft gauge fields correspond to extended field
configurations (shaded) with a typical size $R$ of order $(g^2
T)^{-1}$. Since the soft fields are changing in time, they are
associated with a long wavelength electric field.  The hard modes
act like almost free particles moving on light-like trajectories
(thick lines).  They absorb energy from the long wavelength electric
field which leads to the damping of the soft dynamics. The
semi-hard field modes (thin lines) are responsible for color changing
small angle scattering of the hard particles. In the leading log
approximation, the typical distance between these scattering events is
small compared with $R$, so that the damping becomes effectively local.}
\mlabel{fig.scattering}
\end{figure}

It is remarkable that within the present approximation the effect of
the high momentum ($k\gg g^2 T$) modes can be simply described by a
local damping term and a Gaussian white noise. The hard modes affect
the soft dynamics by Landau damping which, in general, is a non-local
effect.  The modes with $g^2 T \log(1/g) \lsim k\lsim g T$ are
responsible for interactions between the hard particles such that
their mean free path is of order $(g^2 T\log(1/g))^{-1}$. In the limit
of very weak coupling, this mean free path is much smaller than the
size $R\sim (g^2 T)^{-1}$ of a soft gauge field
configuration. Consequently, the soft modes are damped on a relatively
small length scale which means that the damping is effectively local
(see Fig.~\ref{fig.scattering}). This effect is characteristic for
non-Abelian theories, even though the mean free path (or lifetime) of
hard particles has the same parametric form in Abelian and non-Abelian
theories \cite{free.path}. However, the mean free path is determined
by small angle scattering which hardly changes the momenta of the hard
particles. It is the exchange of color charge which makes the damping
local (see also the discussion at the end of
\kapitel{sec.log.boltzmann}).

\subsection{The characteristic amplitudes and frequencies 
of the soft fields revisited} \mlabel{sec.log.scales}
Now that we know the effective theory for the soft modes, at least at
leading logarithmic order, we can complete the estimates made in
\kapitel{sec.scales}. As in \kapitel{sec.scales}, we will not be
concerned about logarithms. \eq{timescale} shows that the estimate
\mref{g4t} is indeed correct. From this, one immediately obtains the
estimate for the amplitude of $A(P)$ in \eq{estimate.A}.
The amplitude of $W(P,\vv)$ can be estimated from the results of
\kapitel{sec.solving.boltzmann},
\begin{eqnarray}
	W(P,\vv) \sim g^{-2} T^{-1} \xi_0(P,\vv) 
	\sim g^{-7} T^{-3}
	,
\end{eqnarray}
where \eq{F.31.5} has been used in the second step.
 
\msection{The hot sphaleron rate} \mlabel{sec.sphaleron}
The rate for electroweak baryon number violation is proportional to
the number of topology changing transitions per unit time and unit
volume \cite{rubakov}.  These transitions are unsuppressed only if the
participating gauge fields have a typical spatial momentum of order
$g^2 T$ \cite{arnoldmclerran}. Furthermore, they require that the
gauge fields change by an amount as large as the field itself, $\delta
A\ \sim A$.  This is the dynamics which is described by Eqs.\
(\ref{langevin2}) and (\ref{langevintag}).

Since this effective theory has no dependence on the UV cutoff
\cite{asy2}, there is only one length scale $R\sim(g^2 T)^{-1}$, and
only one time scale $t\sim (g^4 \log(1/g)T)^{-1}$ left in the
problem. Consequently, the hot sphaleron rate can be estimated
as\footnote{Again, this estimate does not rely on perturbation
theory.}
\begin{eqnarray}
	\Gamma \sim \frac{1}{t R^3} \sim g^{10} \log(1/g) T^4
	.
\end{eqnarray}
Therefore, at leading logarithmic order, the hot sphaleron rate has
the form
\begin{eqnarray}
	\mlabel{rate.2}
  \Gamma = \kappa g^{10}\log(1/g)T^{4}
	,
\end{eqnarray}
where $\kappa$ is a non-perturbative coefficient which does not depend
on the gauge coupling and which has been determined by solving 
\mref{langevintag} on the lattice by Moore \cite{moorelog}.

{\bf Acknowledgements.} I am grateful to Peter Arnold, Guy Moore,
Bert-Jan Nauta, Tomislav Prokopec, Kari Rummukainen, Dam Son, and
Larry Yaffe for useful discussions, and to Mikko Laine, and
Benoit Vanderheyden for critical comments on the manuscript.  I would
also like to thank the organizers and participants of the ITP
workshop {\em Non-equilibrium quantum fields}, Santa Barbara (January
1999), where this work was presented, for a very stimulating
atmosphere.  This work was supported in part by the TMR network
``Finite temperature phase transitions in particle physics'', EU
contract no. ERBFMRXCT97-0122.

\appendix
\section*{Appendix A}
\renewcommand{\theequation}{A.\arabic{equation}}\setcounter{equation}0
For this paper to be self-contained, this Appendix contains some basic
properties of the Laplace transformation \mref{laplace} which are used
in the main text. The original function $f(t)$ is recovered by the
inverse transformation
\begin{eqnarray}
        f(t) = 
        \,\,\,
        \intsub{\im k_0 > 0}
        \,\,\,\,\,
         \frac{dk_0}{2\pi} e^{-ik_0} f(k_0)
	,
\end{eqnarray}
where the integration contour goes from $-\infty + i\times\im k_0$ to
$\infty + i\times\im k_0$. The Laplace transform of a product is
\begin{eqnarray}
	\mlabel{laplace.product}        
        \int_0^\infty dt \, e^{i k_0 t} f(t) g(t) = 
        \intsub{0<\im p_0<\im k_0}
        \frac{dp_0}{2\pi} f(p_0) g(k_0 - p_0)
	.
\end{eqnarray}
To Laplace transform equations of motion one needs the
relations
\begin{eqnarray}
        \mlabel{derivative.1}
        \int_0^\infty dt e^{i k_0 t} \dot{f}(t) &=& - f_\in - i k_0 f(k_0)
	,
\\
        \mlabel{derivative.2}
        \int_0^\infty dt e^{i k_0 t} \ddot{f}(t) &=& 
        - \dot{f}_\in + i k_0 f_\in - k_0^2 f(k_0) 
	,
\end{eqnarray}
where the subscript ``in'' denotes the initial values at $t=0$.
\section*{Appendix B}
\renewcommand{\theequation}{B.\arabic{equation}}\setcounter{equation}0
In the calculation of $\lav\xi_1\rav$ and of the 2-point function of
the noise $\xi_0$ in \kapitel{sec.logarithmic} one encounters
correlation functions of the transverse free semi-hard fields. This
Appendix describes how these correlation functions are obtained from
the solutions to the free equations of motion \mref{solutiona},
\mref{solutionw}, and the Hamiltonian \mref{hamiltonian}.
 
The correlation functions of the initial values $\va_\in$ and $w_\in$
are obtained by thermal average using the Hamiltonian
\mref{hamiltonian}. Since we are only interested in lowest order
perturbation theory, the Hamiltonian can be linearized. The 2-point
function of the transverse gauge fields is the 3-dimensional gauge
field propagator
\begin{eqnarray}
        \mlabel{ainain}
        \langle a^{ia}_\in(\vk_1) a^{jb}_\in(\vk_2) \rangle &=& 
        \projector_\tr^{ij}(\vk_1) \frac{T}{k_1^2} 
        \delta^{ab}(2\pi)^3\delta^3(\vk_1 + \vk_2)
	.
\end{eqnarray}
Evaluating the 2-point function of $w_\in$, one has to account for the
Gau{\ss} law. It can be ignored for the leading logarithmic
approximation in \kapitel{sec.logarithmic}, where only the
transverse semi-hard fields contribute. Then, one can simply use
\begin{eqnarray}
        \mlabel{F.12.3.2}
        \langle w_\in^a(\vk_1,\vv_1)w_\in^b(\vk_2,\vv_2) \rangle &=&
        \delta^{ab}  \frac{T}{\mmdebye} 
        (2\pi)^3\delta^3(\vk_1 + \vk_2)
	4\pi\deltav(\vv_1 - \vv_2) 	
	.
\end{eqnarray}
Using Eqs.\ \mref{solutiona} and \mref{solutionw} together with
\mref{ainain} and \mref{F.12.3.2} one obtains
\begin{eqnarray}
        \mlabel{aain}
        \langle a^{ia}_0(K_1) a^{jb}_\in(\vk_2) \rangle &=& 
        -i\frac{T}{k_1^0} 
        \Big[\stern\Delta_{11}^{ij}(K) - \stern\Delta_{11}^{ij}(0,\vk) \Big]
        \delta^{ab}(2\pi)^3\delta^3(\vk_1 + \vk_2)
	,
\end{eqnarray}
\begin{eqnarray}
        \mlabel{wwin}
        \langle w^a_0(K_1,\vv_1)w_\in^b(\vk_2,\vv_2) \rangle &=&
        \frac{T}{\mmdebye} \Delta_{22}(K_1,\vv_1,\vv_2)
        \delta^{ab}
        (2\pi)^3\delta^3(\vk_1 + \vk_2)
	,
\end{eqnarray}
\begin{eqnarray}
        \langle a^{ia}_0(K_1)w_\in^b(\vk_2,\vv)\rangle =
        \langle w^a_0(K_1,\vv) a^{ib}_\in(\vk_2)\rangle
        =~~~~~~~~~~~~~~~~~~~~~&&\nn
\\
        \mlabel{awin}
        \frac{T}{\mmdebye} \Delta_{12}^i(K_1,\vv)
        \delta^{ab}
        (2\pi)^3\delta^3(\vk_1 + \vk_2)
	,
\end{eqnarray}
where the propagators $\Delta_{\alpha\beta}$ are given by
\eqs{delta11}-\mref{delta22}.

To compute correlation functions of $\va_0$ and $w_0$, it is convenient
to use translation invariance in time to write
\begin{eqnarray}
        \langle a^{ia}_0(x_1) a^{jb}_0(x_2)\rangle
        &=& 
        \Theta(t_1-t_2)
        \langle a^{ia}_0(t_1 - t_2,\vx_1) a^{jb}_\in(\vx_2)\rangle \nn
\\
        \mlabel{F.13.1}
        &&{}+
        \Theta(t_2-t_1)
        \langle a^{ia}_\in(\vx_1)a^{jb}_0(t_2 - t_1,\vx_2) \rangle 
	.
\end{eqnarray}
Taking the Laplace transform, this becomes 
\begin{eqnarray}
        \mlabel{F.13.2}
        \langle a^{ia}_0(K_1) a^{jb}_0(K_2)\rangle =
        \frac{i}{k_1^0 + k_2^0} 
        \[\langle a^{ia}_0(K_1)a^{jb}_\in(\vk_2)\rangle
        + \langle a^{ia}_\in(\vk_1)a^{jb}_0(K_2)\rangle\]
	.
\end{eqnarray}
For Laplace transformed functions the factor
\begin{eqnarray}
  \mlabel{laplace.delta}
  \frac{i}{k_1^0 + k_2^0}
\end{eqnarray}
is the analogue of the delta-function
$2\pi \delta(k_1^0 + k_2^0)$ in Fourier space (cf. \eq{xi.7.1}).
Combining Eqs.~(B.3) and (B.7) we find
\begin{eqnarray}
        \langle a_0^{ia}(K_1) a_0^{jb}(K_2) \rangle
         = -T \delta^{ab} \frac{i}{k_1^0 + k_2^0}
        (2\pi)^3 \delta^3 (\vk_1 + \vk_2)
        ~~~~~~~~~~~~~~~~~~
        &&\nn 
\\
        \mlabel{aa}
        \[ \frac{i}{k_1^0}
        \( \Delta_{11}^{ij}(K_1) - \Delta_{11}^{ij}(0,\vk_1)\)
        + \frac{i}{k_2^0}
        \( \Delta_{11}^{ij}(K_2) - \Delta_{11}^{ij}(0,\vk_2)\) \]&&
	.
\end{eqnarray}
Proceeding similarly for $w_0$ one obtains
\begin{eqnarray}
        \langle a_0^{ia}(K_1) w_0^{b}(K_2,\vv) \rangle
         &=& \frac{T}{\mmdebye} \delta^{ab} \frac{i}{k_1^0 + k_2^0}
        (2\pi)^3 \delta^3 (\vk_1 + \vk_2)\nn \\
        \mlabel{aw}
        && \hspace{1cm} 
        \[  \Delta_{12}^{i}(K_1,\vv) + \Delta_{12}^{i}(K_2,\vv) \]
	,
\end{eqnarray}
and
\begin{eqnarray}
        \langle w_0^{a}(K_1,\vv_1) w_0^{b}(K_2,\vv_2) \rangle
         &=& \frac{T}{\mmdebye} \delta^{ab} \frac{i}{k_1^0 + k_2^0}
        (2\pi)^3 \delta^3 (\vk_1 + \vk_2)\nn \\
        \mlabel{ww}
        && 
        \Big[  \Delta_{22}(K_1,\vv_1,\vv_2) 
        + \Delta_{22}(K_2,\vv_1,\vv_2) \Big]
	.
\end{eqnarray}
\section*{Appendix C}
\renewcommand{\theequation}{C.\arabic{equation}}\setcounter{equation}0
In \kapitel{sec.xiexpansion} it was argued that terms proportional to
$A^n$ do not occur in $\lav\xi_n\rav$, because they would violate
gauge covariance. However, if one writes down the perturbative
expansion \mref{xiexpansion} for $\xi$, one encounters terms containing
$A$. It is interesting to see how these terms cancel after the
replacement~\mref{replacechi}.  

Consider $h$ in \mref{F.1.1.5} and keep the term $A$.  The
corresponding contribution to $\xi_1$ is given by \mref{xi1}, where
$a_1$ and $w_1$ are now linear in $A$. Using \mref{F.16.2},
\mref{F.16.3} one obtains
\begin{eqnarray}
        \[\xi_1^a(P,\vv)\]_A &=& - g^2 f^{abc} 
        \!\!
        \intsub{0<\im k^0<\im p^0 }
        \!\!
        \frac{d^4 k}{(2\pi)^4}
        \intv{1} 
        \!\!
        \intsub{0<\im p_1^0<\im k^0 }
        \!\!
        \frac{d^4 p_1}{(2\pi)^4} v_1\cdot A^d(P_1)
\nn\\ 
        && {}  \[
        f^{bde} v^i \Delta_{12}^{i}(K,\vv_1) 
        w_0^e(K-P_1,\vv_1)  w_0^c(P - K,\vv)        \right. \nn\\
        \mlabel{xi1.a}
        && 
        \left.
        {}+  f^{cde}  \Delta_{22}(K,\vv,\vv_1) 
        \vv\cdot\va_0^b(P-K) w_0^e(K-P_1,\vv_1)\] 
	.
\end{eqnarray}
Now we insert the expressions for $\va_0$ and $w_0$ which were
obtained in \kapitel{sec.solving.eom} and replace the products of the
free semi-hard fields by their thermal expectation values according
to~\mref{replacechi}.  Using \eqs{aw}, \mref{ww} we obtain
\begin{eqnarray}
\lefteqn{
  \lav\xi_1^a(P,\vv)\rav_A  = \frac{N g^2 T  }{\mmdebye}
        \!\!
        \intsub{0<\im k^0<\im p^0 }
        \!\!
        \frac{d^4 k}{(2\pi)^4}
        \intv{1} 
        \!\!
        \intsub{0<\im p_1^0<\im k^0 }
        \!\!
        \frac{d^4 p_1}{(2\pi)^4}  \frac{i}{p^0 - p_1^0}
        (2\pi)^3 \delta^3 (\vp - \vp_1)
         }\nn
\\ 
        && v_1\mal A^a(P_1) \Big\{     
        v^i \Delta_{12}^{i}(K,\vv_1) 
        \Big[\Delta_{22}(K-P_1,\vv,\vv_1) + \Delta_{22}(P-K,\vv,\vv_1) \Big]
        \nn
\\ && 
        \mlabel{xi.8.2}
        \phantom{\mbox{$ppppp$}} 
        {} - \Delta_{22}(K,\vv,\vv_1)v^i 
        \[ 
         \Delta_{12}^{i}(P-K,\vv_1) + \Delta_{12}^{i}(K-P_1,\vv_1)
        \right]
        \Big\} 
	.
\end{eqnarray}
As in \eq{xi1replace1}, the terms in the curly bracket which do not
depend on $P$ vanish and one obtains
\begin{eqnarray}
  \mlabel{xi.9.1}
  &&
  \lav\xi_1^a(P,\vv)\rav_A  = \frac{N g^2 T  }{\mmdebye}
        \!\!
        \intsub{0<\im k^0<\im p^0 }
        \!\!
        \frac{d^4 k}{(2\pi)^4}
        \intv{1}v_1\mal A^a(P)  
\\
  &&\nn
  \[
     v^i \Delta_{12}^{i}(K,\vv_1)\Delta_{22}(P-K,\vv,\vv_1)
     - \Delta_{22}(K,\vv,\vv_1)v^i \Delta_{12}^{i}(P-K,\vv_1)
  \] .
\end{eqnarray}
The two terms in the square bracket cancel after the substitution
$K\to P-K$ in the second term. Thus, we find that indeed
\begin{eqnarray}
  \mlabel{xi.9.2} \lav\xi_1^a(P,\vv)\rav_A =0
	.
\end{eqnarray}

A similar cancellation (I do not write out the straightforward
calculation explicitly, instead I am restricting myself to a qualitative
description) takes place to the $A$-dependent terms in
\begin{eqnarray}
        \xi_2^a(P,\vv) &=& g f^{abc} 
	\intsub{0<\im k^0 < \im p^0}
        \frac{d^4 k}{(2\pi)^4} 
	\Big[ \vv\cdot \va_2^b(K) w_0^c(P -
        K,\vv) 
\nn \\ 
	\mlabel{xi.10.1} 
	&& {}+ \vv\cdot \va_1^b(K)
        w_1^c(P - K,\vv) + \vv\cdot \va_0^b(K) w_2^c(P - K,\vv) \Big]
	.
\end{eqnarray}
Making repeated use of \mref{F.16.2} and \mref{F.16.3}, one can write
$\va_n$ and $w_n$ for $n=1,2$ in terms of $\va_0$, $w_0$, and
$A$. After the replacement \mref{replacechi} none of the three terms
in \mref{xi.10.1} vanishes individually. Each of them consists of two
parts. The first part of the first term vanishes because the
integration contour can be moved to infinity. The second part cancels
the first part of the second term. The second part of the second term
cancels the first part of the third term. Finally, in the second part
of the third term the integration contour can be moved to infinity, so
that it does not contribute.

\section*{Appendix D}
\renewcommand{\theequation}{D.\arabic{equation}}\setcounter{equation}0
In this Appendix we determine the eigenvalues of the collision term in
\mref{boltzmann.2}. The $l=1$ eigenvalue enters the damping
coefficient in the Langevin equation \mref{langevintag}. Furthermore,
consistency of the solution \mref{wi.solution} to the $l=1$ Boltzmann
equation \mref{boltzmann.wi} requires the $l\ge 2$ eigenvalues to be
non-zero (see \eq{lm.solution}).

Since the integral kernel of the collision term is proportional to
$I(\vv,\vv')$ in \eq{kern}, we consider the eigenvalue equation
\begin{eqnarray}
         \int d\Omega_{\vv'}I(\vv,\vv') Y_{lm}(\vv') 
        = \lambda_l Y_{lm}(\vv)
	.
\end{eqnarray}
\eq{vanish} implies that $\lambda_0 = 0$, which is necessary for the
current \mref{current} to be conserved.  To determine $\lambda_l$ for
$l\ge 1$ it is sufficient to consider the eigenvalue equation for
$Y_{l0}(\cos\theta)$, which is proportional to the Legendre polynomial
$P_l(\cos\theta)$. Thus the eigenvalues of the kernel $I$ are
determined by
\begin{eqnarray}
	\int d\Omega_{\vv'}
	I(\vv,\vv') P_l(\hat{\bf z}\mal\vv') = 
        \lambda_l P_l(\hat{\bf z}\mal\vv)
	.
\end{eqnarray}
It is also sufficient to consider one special value for $\vv$, a
convenient choice is $\vv = \hat{\bf z}$. Then, we simply have
$P_l(\hat{\bf z}\mal\vv) =P_l(1) = 1$, so that
\begin{eqnarray}
        \mlabel{Y.2.2.3}
                \lambda_l = 
        -1 + \frac2\pi\int\limits_{-1}^1 d x \frac{x^2}{\sqrt{1-x^2}}P_l(x)
	.
\end{eqnarray}
For odd $l$ we have $\lambda_l = -1$, because in this case the second
term in \mref{Y.2.2.3} vanishes. For even $l\ge 2$ the integral in
\mref{Y.2.2.3} is always smaller than 1, due to $P_l(x)\le 1$.  Thus
all $\lambda_l$ are non-zero and negative when $l\ge 1$.

\end{document}